\documentclass[a4paper,11pt]{article}
\pdfoutput=1
\usepackage{jheppub} 
\usepackage[T1]{fontenc} 
\usepackage{amsmath,amssymb,graphicx,bbm,mathrsfs,nicefrac}
\usepackage{tikz}

\newcommand{\be}{\begin{equation}} 
\newcommand{\ee}{\end{equation}}  
\newcommand{\bea}{\begin{eqnarray}}  
\newcommand{\eea}{\end{eqnarray}}  
\def\bpm{\begin{pmatrix}}
\def\epm{\end{pmatrix}}

\def\gev{\, {\rm GeV}}
\def\tev{\, {\rm TeV}}

\def\bsp#1\esp{\begin{split}#1\end{split}}

\newcommand\lsim{\mathrel{\rlap{\lower4pt\hbox{\hskip1pt$\sim$}}
    \raise1pt\hbox{$<$}}}
\newcommand\gsim{\mathrel{\rlap{\lower4pt\hbox{\hskip1pt$\sim$}}
    \raise1pt\hbox{$>$}}}

\newcommand{\captionfonts}{\small}

\newcommand{\approptoinn}[2]{\mathrel{\vcenter{
  \offinterlineskip\halign{\hfil$##$\cr
    #1\propto\cr\noalign{\kern2pt}#1\sim\cr\noalign{\kern-2pt}}}}}

\makeatletter  % Allow the use of @ in command names
\long\def\@akecaption#1#2{%
  \vskip\abovecaptionskip
  \sbox\@tempboxa{{\captionfonts #1: #2}}%
  \ifdim \wd\@tempboxa >\hsize
    {\captionfonts #1: #2\par}
  \else
    \hbox to\hsize{\hfil\box\@tempboxa\hfil}%
  \fi
  \vskip\belowcaptionskip}
  
\makeatother   % Cancel the effect of \makeatletter

\preprint{CERN-PH-TH/2013-199}
\title{Probing top-partners in Higgs + jets}

\author[a]{Andrea Banfi}
\author[b,c]{, Adam Martin}
\author[a]{and Ver\'onica Sanz}
\affiliation[a]{Department of Physics and Astronomy, University of Sussex, Brighton BN1 9QH, UK}
\affiliation[b]{Department of Physics, University of Notre Dame, Notre Dame, IN 46556, USA}
\affiliation[c]{Theory Division, Physics Department, CERN, CH-1211 Geneva 23,
  Switzerland}

\abstract{
Fermionic top-partners arise in models such as Composite Higgs and Little Higgs. They modify Higgs properties, in particular how the Higgs couples to top quarks. Alas, there is a low-energy  cancellation acting in the coupling of the Higgs boson to gluons and photons.  As a result of this cancellation, no information about the spectrum and couplings of the top-partners can be obtained in $g g \to h$, just the overall new physics scale $f$. In this paper we show that this is not the case when hard radiation is taken into account. Indeed, differential distributions in Higgs plus jets are sensitive to the top-partner mass and coupling to the Higgs. We exploit the transverse momentum distribution of the hard jet to obtain limits on the top-partners in the $14 \tev$ LHC run, finding that 300 fb$^{-1}$ of data of $14\,\tev$ LHC are sufficient to rule out top-sector mixing angles $\sin^2 (\theta_R)$ > 0.05 for  top-partners with masses from 300 GeV to above 2 TeV.
}
\emailAdd{a.banfi@sussex.ac.uk}
\emailAdd{adam.martin@cern.ch}
\emailAdd{v.sanz@sussex.ac.uk}


\begin{document}
\maketitle
\flushbottom

\section{Introduction}

The idea that the Higgs is a composite resonance, manifestation of the breaking at high energies of a global symmetry is an old one~\cite{HasPGB}. This idea has been thoroughly explored and explicit realizations are built as Little Higgs (LH)~\cite{LHs}, Composite Higgs (CH)~\cite{CHMs} and Partial compositeness~\cite{partial-comp} models.
In these models, the pseudo-Goldstone nature of the Higgs explains why other resonances of the new sector have not been seen yet, but introduces the hurdle of how to generate a potential for the Higgs, a mass and self-interactions. Successful electroweak symmetry breaking requires new states to generate a sizable potential, and those are typically top-partners. Top-partners are heavy resonances with the same quantum numbers as the top and couple strongly to the Higgs. Their contribution is essential to raise the Higgs mass to acceptable levels. 

Top-partners are then a key piece to understand electroweak symmetry breaking, but searching for them is more complicated that one would expect. Although they contribute to the $hgg$ coupling, there is a cancellation at low energies which renders this coupling insensitive to the mass and coupling of the top-partner~\cite{cancellation,cancellation-nnlo}. Instead, the coupling is only sensitive to $v^2/f^2$, where $f$ is the scale of breaking of the global symmetry leading to the pseudo-Goldstone sector.  As a result, fits on the rates of Higgs production and decay into various final states are only sensitive to this parameter~\cite{others-higgs,fits-comp}, and not to the individual coupling and mass of the top-partners.  Double Higgs production $pp \to h h$~\cite{doubleH} is one obvious place to look for signs of  top-partners. However, this process has a small cross section that is largely insensitive to finite top-partner masses. 

Top-partners can be searched for directly, both produced in pairs $pp \to T\bar T$ and singly produced $pp\to T + X$. However, direct production carries more model dependence since the search strategies and limits depend on how the top-partner decays -- an aspect of the model usually unrelated to electroweak symmetry breaking. Most experimental searches focus on the decay modes $T \to W^+ b, T \to Z t$, and/or $T \to h t$. The bounds, assuming these three modes are the only ones available are roughly $700-800\, \gev$~\cite{exp-vectorlike}. However if there are other decays possible, such as to exotic pseudo-Goldstone states~\cite{Kearney:2013cca,Kearney:2013oia}, the $T$ width will increase and the bounds will weaken. 

Associated top-partner plus Higgs production $pp \to T \bar T h$ is one way to directly test the $hTT$ coupling. However, the cross section for this process falls steeply as the top-partners become heavy. Additionally, this method requires reconstructing the $T$ produced with the Higgs, which is sensitive to the model-dependent details of how $T$ decays. 

\begin{figure*}[h!]
\centering
\includegraphics[width=0.45\textwidth]{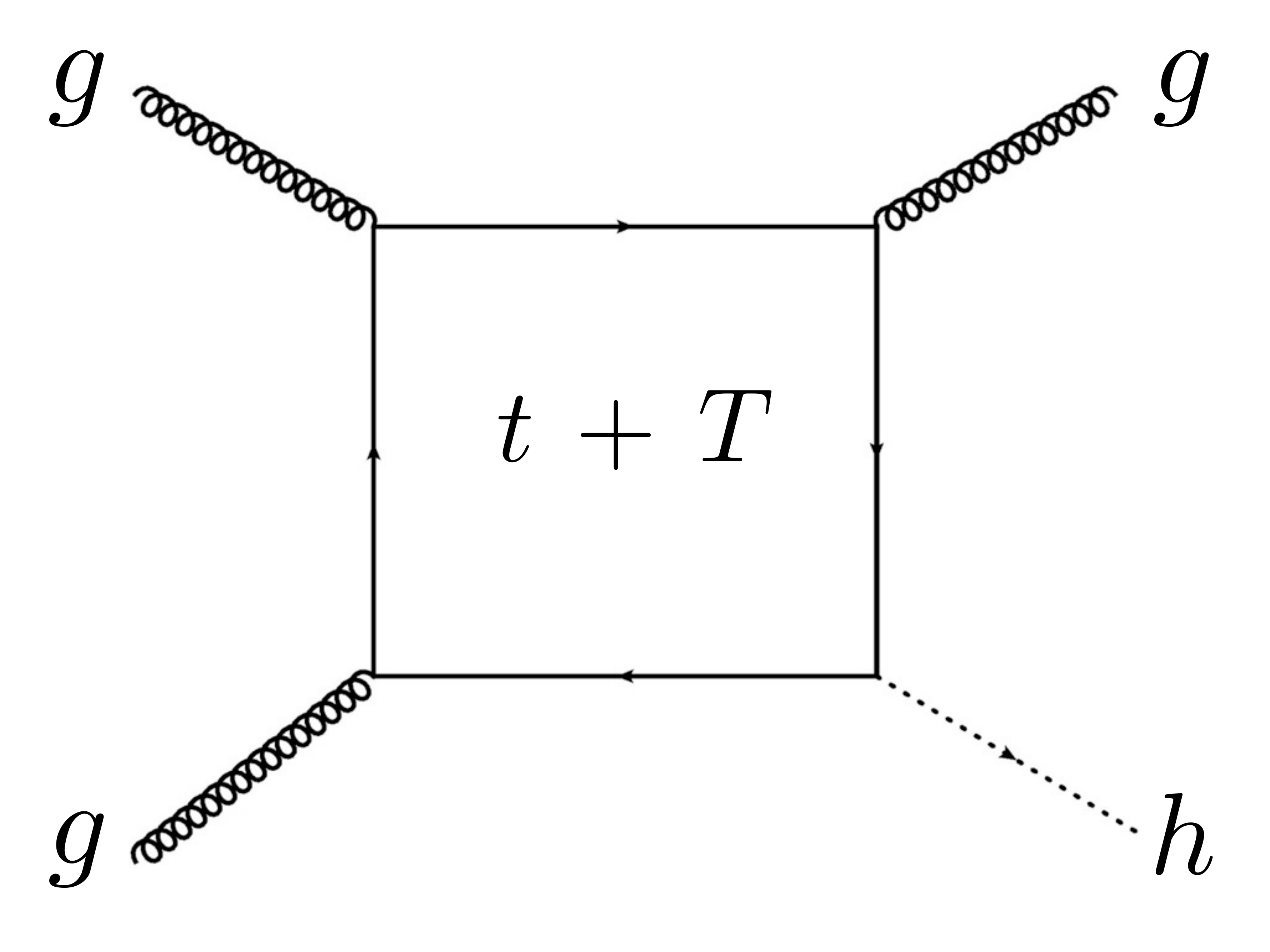}
\includegraphics[width=0.45\textwidth]{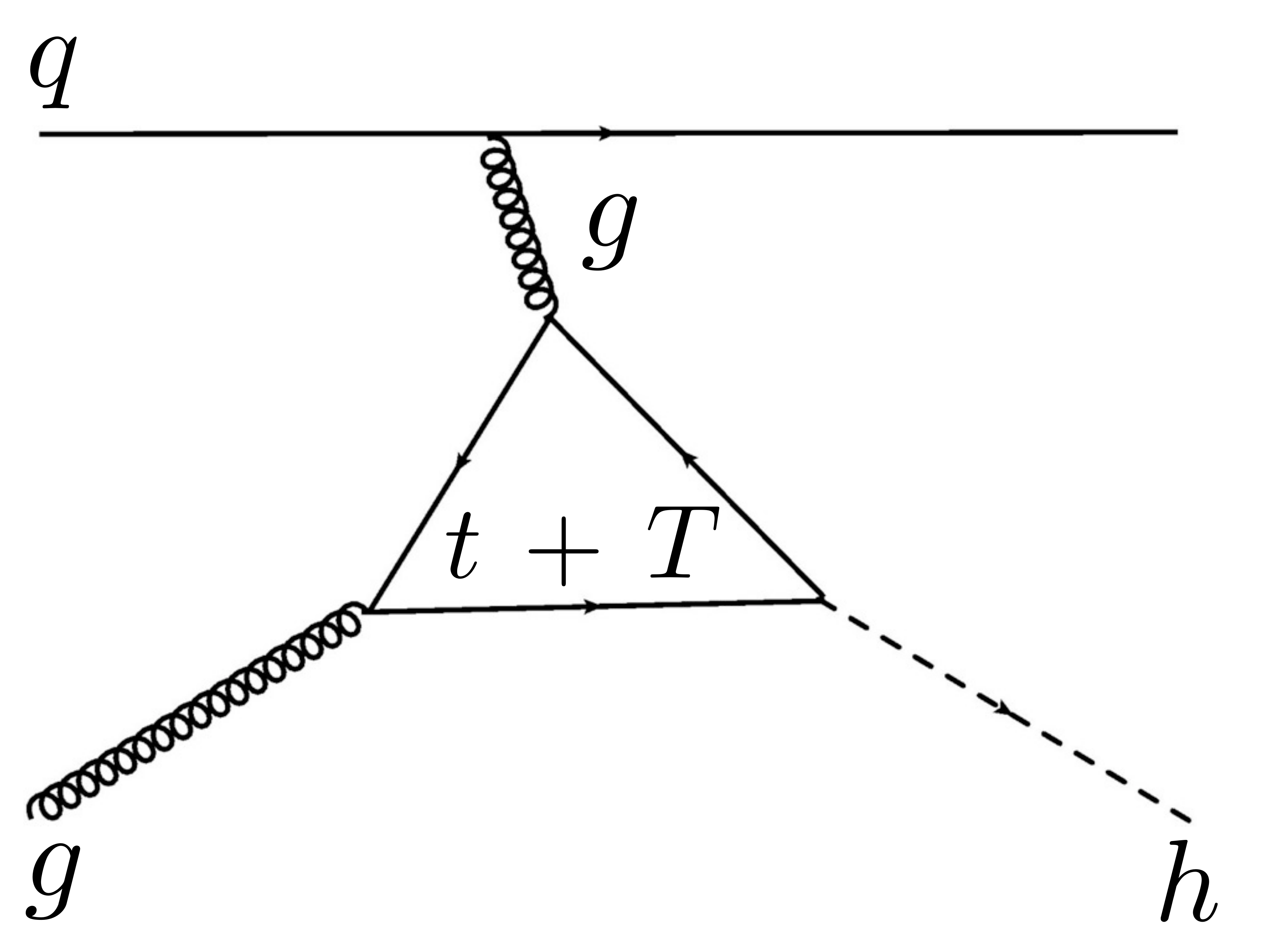}
\caption{\it Typical diagrams contributing to $p p \to h + j$.}
\label{fig:feyn}
\end{figure*}

In this paper we show that, unlike $pp \to h$, the process $pp \to$ Higgs plus high-$p_T$ jet is sensitive to the individual coupling and mass of the top-partner. The reason for this can by seen by inspecting the diagrams that contribute to $pp \to h+j$. One contribution comes from box diagrams, shown on the left in Fig.~\ref{fig:feyn}. As the additional gluon probes the fermion loop, it is not surprising that these diagrams carry a dependence on the fermion mass. The second contribution to $pp \to h +j$ comes from familiar $hgg$ triangle diagrams stitched on to additional partons. Because of their similarity to $gg \to h$ diagrams, one may think these diagrams are not sensitive to the internal fermion mass. This is not true; to make a final state with high-$p_T$, the intermediate gluon in diagrams such as the right side of Fig.~\ref{fig:feyn} must have high virtuality. The high virtuality of the incoming gluon means the fermion triangle is resolved at a different, shorter scale compared to $gg \to h$ production, and the process becomes sensitive to the fermion mass. Therefore, by studying $pp \to h+j$ and comparing to SM rates, one can bound the top-partner and its Higgs coupling independent of the details of the $T$ decay.

The setup for the remainder of this paper is as follows: In Sec.~\ref{top-part}, we describe the low-energy cancellation at the level of dimension-six operators, and how the extra jet would come from an effective theory including dimension-eight operators. We also set the notation and translation between our parametrization and common models in the literature. In Sec.~\ref{Hj}, we show that the sensitivity to mass and couplings arise as double logarithmic terms in the matrix element in the high-$p_T$ limit. We then numerically study $h+j$ production, both at parton level and after including parton distribution functions. In Sec.~\ref{HOC} we discuss the stability of our results at leading
order when higher order corrections and experimental uncertainties would be included. Finally, in Sec.~\ref{seclimits}, we use the
differential distributions to set limits on the top-partner masses and couplings in the $14\,\tev$ LHC run.

\section{Top-partners}~\label{top-part}

In this section we describe the effects on Higgs production due to a new colored fermion that mixes with the top quark, which we will call the {\it top-partner}. To set limits, we present a explicit choice of mass mixing, which can be mapped into from several CH and LH models. Our study will focus on the Higgs production in association with jets, and, in the next section, we explain why one requires extra hard radiation. 

% but note that the method we propose here is independent of the explicit choice of mass matrix{\bf ???}. 
\subsection{Mass matrix}

In this section we parametrize the top-partner sector as a Dirac fermion $T=(T_L,T_R)$, with mass $M$ and a mixing with the SM top given by $\Delta$.
Without loss of generality, one can then write the mass matrix between the top $t_{L,R}$ and top-partner $T_{L,R} $ as 
\begin{equation}
(\bar t_L\, \bar T_L)\left( \begin{array}{cc} \frac{y_t h}{\sqrt 2} & \Delta \\ 0 & M \end{array} \right)_{h = v}\left(\begin{array}{c} t_R \\ T_R \end{array} \right).
\label{themassM}
\end{equation}

This mass matrix can be diagonalized by a bi-unitary transformation, on the left with a mixing angle of $\theta_L$ and on the right with a mixing angle of $\theta_R$. Identifying $\frac{y_t v}{\sqrt 2} = m$, we can trade the three parameters $m, \Delta, M$ for the two mass eigenstates $m_t, M_T$ and one of the mixing angles $\theta_R$. Expanding the Higgs about its vacuum expectation value, we can then find the couplings of the mass eigenstate top-quark/top-partner to the Higgs boson. The diagonal Higgs couplings in terms of $m_t, M_T, \theta_R$ are:
\be
h\, \overline{t} t :  \frac{m_t}{v}\cos^2({\theta_R}),\quad h\, \overline{T} T:  \frac{M_T}{v}\sin^2({\theta_R}) \ ,
\label{eq:hcoupl}
\ee
where 
\bea
\theta_R = \frac{1}{2} \arcsin\left(\frac{2 m_t M_T \eta}{M_T^2-m_t^2} \right),\quad \tan{\theta_L} = \frac{M_T}{m_t}\tan{\theta_R}
\label{couptTb}
\eea
and $\eta=\Delta/M$. 

From the coupling expressions in Eq.~(\ref{eq:hcoupl}), we can quickly see that the $gg \to h$ amplitude is insensitive to the mixing angle $\theta_R$. The $gg \to h$ amplitude from a single fermion loop can be written as~\cite{ggh-ref}:
\be
A_i (gg \to h) = \frac{\alpha_s m^2_H}{4\pi v}\, \kappa\, \Big(\frac{2 - 4\,m^2_{f,i}\,(1-\tau) C_0(4\,m^2_{f,i} \tau; m^2_{f,i})}{\tau} \Big) \equiv \frac{\alpha_s m^2_H}{4\pi v}\, \kappa\, \mathcal A_i\,,
\ee
where $ \tau = \frac{m^2_H}{4m^2_{f,i}}, \kappa =
y_{f,i}\Big(\frac{v}{m_{f,i}} \Big)$ and $C_0$ is the three-point
Passarino-Veltman function, see Ref.~\cite{ggh-ref} for conventions and the explicit form of $C_0$. When the fermion running in the loop is heavy ($\tau \to 0$), $\mathcal A_i$ asymptotes to a constant value, $\mathcal A(0) = -4/3$ and the amplitude is independent of the fermion mass. Therefore, if we combine two $gg \to h$ amplitudes, {\em both} coming from fermions far heavier than the Higgs mass, the net amplitude also insensitive to the individual fermion masses and only depends on the strength of the Higgs-fermion couplings:
\be
A_t(gg \to h) + A_T(gg \to h)|_{m_h \ll 2m_t, 2M_T} = \frac{\alpha_s m^2_H}{4\pi v}\Big(-\frac{4}{3} \Big) (\kappa_{t} + \kappa_{T} )\,.
\label{eq:heavy}
\ee
Plugging in the couplings in in Eq.~(\ref{eq:hcoupl}), we find the $gg \to h$ amplitude is independent of the mixing between the top and top-partner up to corrections of $O\Big(\frac{m^2_H}{4m^2_t} \Big)$.

A crucial requirement for this insensitivity is that both fermions are heavy compared to the external momenta. This requirement teaches us two things. First, it tells us that this insensitivity of $gg \to h$ to SM fermion - new fermion mixing is only possible for the top sector; all other quarks are light compared to $m^2_H$ so Eq.~(\ref{eq:heavy}) no longer holds\footnote{See Ref. \cite{cedric} for constraints on light fermion-new fermion mixing coming from this breakdown.}. Second, being a $2 \to 1$ process, the total invariant mass entering the loop ($\hat s)$ in $gg \to h$ is fixed to $m^2_H$. However, this is not true for more general processes, such as when the Higgs recoils agains other final state particles; there,  $\hat{s} \gg m^2_H$ is possible. We emphasize that, to guarantee $\hat s \gg m^2_H$, one must focus on Higgs production with lots of recoil. Once higher-order corrections to $gg \to h$~\cite{Djouadi:1991tka,Harlander:2002wh,Anastasiou:2005qj,Grazzini:2008tf} are taken into account, the Higgs will acquire some recoil. However, inclusive $pp \to h+X$ is dominated by $p_T \lesssim m_H$, which is insufficient to unveil the properties of the internal fermion loop. We must instead look to Higgses produced in association with one or more high-$p_T$ objects.

%In popular models such as Composite Higgs and Little Higgs, the Higgs is a pseudo-Goldstone boson. This  leads to a very definite prediction in the production of the Higgs (with no hard jets). Due to a low-energy cancellation, the main process involving the top-partner (i.e. gluon fusion) is insensitive to the top-partner mass and coupling.
%
%Before we more onto the details of these models and the cancellation, let us emphasize that the search we propose would provide information on the top-partners, {\it irrespective} of whether the top-partner arises from a model of the Higgs as a pseudo-Goldstone boson or not.

\subsection{Low energy Higgs theorems and the insensitivity of the $hgg$ coupling}
\label{sec:LET}

In this section we describe how a low energy theorem is responsible for the insensitivity of the dimension-five coupling $hgg$.

Consider a colored fermionic particle which transforms under the fundamental of $SU(3)_c$, and whose mass comes at least partially from electroweak symmetry breaking, $M=M(H)$. In this case, it is well known~\cite{Ellis:1975ap, Kniehl:1995tn} that the effect of this particle in the $h g g $ coupling at low energies ($E \ll M$) would be described by
\bea 
{\cal L}_{h^n gg} = \frac{g_s^2}{96\pi^2} G_{\mu\nu}^a G^{a\, \mu\nu}\left( A_{1} h +\frac{1}{2}A_{2}h^{2}+\ldots\right)\,,  
\eea
where the coefficients $A_n$ can be written as
\bea  
A_n \equiv \frac{\partial^{n}}{\partial H^{n}} \ln \det  \mathcal{M}^{\dagger}\mathcal{M} (H) |_{h \to v} \ ,
\eea
$H$ is the Higgs doublet, 
and
$\mathcal{M}$ is the heavy fermion mass matrix. 

In Composite Higgs and in Little Higgs models, the Higgs is a pseudo-Goldstone boson. This property restricts the coupling of the Higgs to fermions, hence the form of $\cal M$. In these models, the form of the mass matrix factorizes as follows
\bea
\det  \mathcal{M}^{\dagger}\mathcal{M} (H) = \rho(H/f) \times \rho'(\textrm{ couplings, masses })
 \label{factor}
\eea
where $f$ is the scale at which the global symmetry is broken, resulting in the appearance of the pseudo-Goldstone boson sector. For example, in the minimal Composite Higgs (i.e. coset $SO(5)/SO(4)$), $\rho=\sin^2(2 H/f)$. This is similar to the fact that the pion non-derivative interactions appear as a function of the spurion $\pi/f_{\pi}$.  As a result of this restriction, when one evaluates the effect of the fermion sector on the $h g g$ coupling, the dependence in the coupling and mass (i.e. the dependence in the piece $\rho'$ in Eq.~\ref{factor}) factors out and one is left with
\bea
\frac{\partial}{\partial H} \ln \det  \mathcal{M}^{\dagger}\mathcal{M} (H) |_{h \to v} =
 \frac{\partial}{\partial H} \ln   \rho(H/f)  |_{h \to v} 
\eea
which is just a function of the parameter
\bea
\xi=\frac{v^2}{f^2}\,.
\eea
All dependence on the coupling and mass of the top-partners disappears in the low energy limit. This statement goes beyond the leading $m \to \infty$ terms calculated in in Eq.~(\ref{eq:heavy}). It tells us that even when higher order $1/m$ terms are taken into account, the $hgg$ coupling in CH and LH models  will only depend on $\xi$.

From the point of view of the effective theory, the inclusion of a hard jet in the final state corresponds to adding higher dimension operators. At the level of processes with one extra gluon, one needs to consider three dimension-seven operators~\cite{dim7} (i.e. dimension-eight operators with one $v$-insertion), which have the form
\bea
h \, \left( c_1 D_{\alpha} \, G_{\mu\nu} D^{\alpha} G^{\mu\nu} + c_2  G_{\nu}^{\mu} G^{\nu}_{\rho} G^{\rho}_{\mu} + c_3 D^{\mu} \, G_{\mu \nu} D_{\alpha} G^{\alpha \nu} \right) \ .
\eea
As we will see in the next Sec.~\ref{Hj}, the effect of top-partners in the processes involving those operators does carry information about the coupling and masses of the top-partners.

\section{Top-partners in pseudo-Goldstone Higgs models}

In models where the Higgs is a pseudo-Goldstone boson and assuming only one top-partner, the coupling of the top ($t_{L,R}$) and top-partner ($T_{L,R}$) mass eigenstates to the Higgs can be written in terms of field-dependent masses:
\bea
-\mathcal{L}_{m}=m_{t}(h)\,\overline{t}_{R}t_{L}+M_{T}(h)\,\overline{T}_{R}T_{L} + \mathrm{h.c.}
\eea
where, at lowest order in the strong scale $f$, $m_t(h)$ and $M_T(h)$ can be parametrized as~\cite{non-LET}
\bea 
m_{t}(h)=\frac{y_{t}h}{\sqrt{2}}\left(1-\frac{c_{t}}{2}\frac{h^{2}}{f^{2}}\right)\,,\qquad M_{T}(h)=\lambda_{T}f\left(1+a_{T}\frac{h^{2}}{f^{2}}\right)\,, \label{mtTa}
\label{field-dep masses}
\eea 
with $a_{T}=\mathcal{O}(y_{t}^{2}/\lambda_{T}^{2})$. The constant $c_t = 2\,a_T + c_{\sigma}$, where $c_{\sigma}$ is a contribution coming from the non-linearity of the Higgs in pseudo-Goldstone models; this piece is model dependent, but $O(\frac{v^2}{f^2})$. Expanding $h \to v + h$ and continuing to work to lowest order in $\xi = v^2/f^2$, the Higgs couplings in Eq.~(\ref{field-dep masses}) can be massaged into the same form as Eq.~(\ref{eq:hcoupl}):
\be
h\, \overline{t} t :  \frac{m_t}{v}( 1- 2 a_T \xi + O(\xi^2)),\quad h\, \overline{T} T:  \frac{M_T}{v} (2 a_T \xi + O( \xi^2) ) ,
\ee
 Therefore, we can identify
 \be
 \sin^2({\theta_R}) = 2 a_T \xi + O(\xi^2),
 \label{eq:CHend}
 \ee
where the $O(\xi^2)$ correction includes the non-linear piece $c_{\sigma}$. 

\begin{figure*}[h!]
\centering
\includegraphics[width=0.45\textwidth]{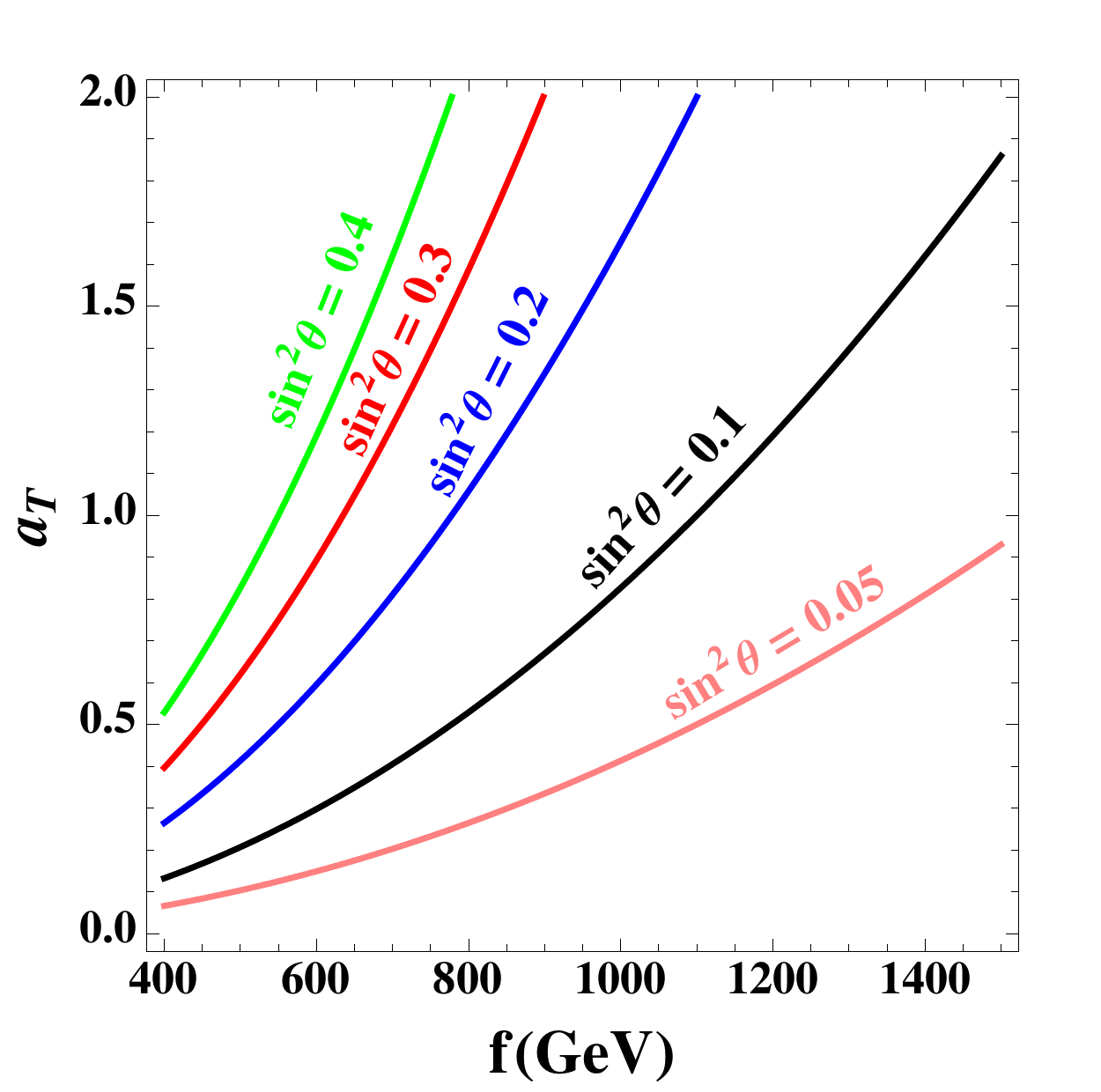}
\caption{\it Lines with constant mixing angle $\sin^2(\theta_R)$ in the ($f, a_T$) plane.}
\label{fig:atf}
\end{figure*}

Despite the fact that CH and LH models come in many varieties and have various field content and underlying symmetry, the mass matrices for the top-partner sector -- at least for several well-studied models -- can all be cast in the form Eq.~(\ref{themassM}) up to terms of order ${\cal O}(1/f^2)$. This mapping is shown explicitly in Appendix~\ref{choices}. Following the steps in Eq.~(\ref{field-dep masses} - \ref{eq:CHend}) for a given CH or LH model, we find
%Although for different models in CH and LH the mass matrices are quite different, see Appendix A, the mapping at order ${\cal O}(1/f^2)$ is the same and reads
\bea
a_T = \frac{c^2 y_t^2}{\lambda_T^2}\,,
\label{eq:whatisa}
\eea
where $c$ is an order one coefficient arising from the linear coupling of elementary and composite fermions. Different CH, LH models yield different $c$. For example, $c=1$ in the littlest Higgs model. In Fig.~\ref{fig:atf} we show the relation between the mixing angle and the parameters in the parametrization in Eq.~(\ref{mtTa}). Large values of $\sin^2(\theta_R)$ imply low values of the scale of breaking of the global symmetry or large coupling $a_T$, i.e. $\lambda_T \simeq y_t$.

\subsection{Current bounds}
\label{bounds}

The scale of breaking $f$ depends on the UV completion of the theory. This scale is subject to electroweak precision tests~\cite{UVvector} and flavor constraints, which depend on the assumptions on the symmetry structure and spectrum of the theory. For example, one could imagine that the UV completion preserves custodial symmetry~\cite{LR}, or that there is a spectrum designed to minimize the $S$ parameter~\cite{S}. One could also assume there is a specific flavor structure~\cite{andi-michele} in the model at the scale $f$ which keeps the flavor constraints under control. Regardless of these UV-sensitive issues, we expect modifications on the way the Higgs realizes electroweak symmetry breaking, hence modifications on the Higgs couplings to SM fields. Keeping an open mind about the UV structure of the top-partner theories, we will consider  $\xi \lesssim 0.3$, the current bounds from Higgs signal-strength fits~\cite{Higgs-LHC} (though the actual bound on $\xi$ depends on the specific model).  In practice, the parameter $\sin^2(\theta_R)$ is more convenient to use than $a_T$ and $\xi$. Motivated by the bounds on $\xi$ and the expression for $a_T$ (Eq.~(\ref{eq:whatisa})), we consider $\sin^2(\theta_R) \le 0.4$ in all numerical studies. 

Searches for top-partners in pair production through color processes, i.e. $pp \to T \, \bar T$ compete with the search we propose here, but the comparison would depend on the electroweak quantum numbers, e.g.~the left and right handed composition~\cite{exp-vectorlike} and what they decay to. The phenomenology could be driven by leptonic channels~\cite{top-hunter} or more complicared multijet or boosted signatures~\cite{andi-me, Kribs:2010ii}. Similarly, single production of the top-partner depends on the flavor structure of the model and how electroweak precision is addressed~\cite{andi-me,matching}.

\section{ The process $p p \rightarrow H + j$}\label{Hj}

Having mapped the top-mixing sector of CH and LH models into our parameterization, we are ready to explore  the effects of top-partners on Higgs plus jet production. We start by looking at some limiting cases, then give numerical results both at parton level and after including the parton distribution functions (PDFs)\footnote{For a very recent study of Higgs plus jets in the context of dimension-seven operators, see Ref.~\cite{Harlander}.}.

\subsection{Generalities}
\label{sec:general}

When the Higgs is produced in association with a jet, the assumptions of the low-energy theorem no longer holds.  Specifically, for a given $p_{T,h} = p_{T,j}$, there is a bound on $\hat s$, $\hat s \ge 2\,p_T \Big( p_T + \sqrt{m^2_H + p^2_T} \Big)$. For sufficient $p_T$, this $\hat s $ is no longer small compared to the mass of the fermion (top, or top-partner) running around the loop, we can no longer take the simple $\hat s \to 0$ limit and must retain the full dependence of the loop functions on  $\hat s/m^2_f$. To get some idea of how the $h + j$ cross section changes with  $\hat s/m^2_f$, we can look at the limiting cases: i.) high-$p_T$ and ii.) low-$p_T$. 

There are four partonic subprocess that contribute to $pp \to h+j$,
\bea
gg \to h+g, gq \to h+q,\, \overline q g \to h + \overline q,\, q \overline q \to h + g \ .
\eea

The actual breakdown of the subprocesses depends on $p_T$, the scale choice, and the PDFs, but  gluon-gluon initiated subprocesses typically dominate, so we focus on $gg \to h + g$ for now. The $gg \to h+g$ cross section can be decomposed as a sum over the various gluon helicity configurations~\cite{crazy, Glover-Baur} and the different fermions running in the loop:
\be
\hat{\sigma}(gg \to gh) = \frac{\beta_H}{16\pi\hat s} \frac{\alpha^3_s}{4\pi\, v^2}\frac 3 2 \Big(\sum\limits_{\lambda_i = \pm} \Big| \sum\limits_{f_i} \mathcal M^i_{\lambda_1\lambda_2\lambda_3}(\hat s, \hat t, \hat u, m_i, y_i) \Big|^2 \Big),
\label{eq:bigM}
\ee
where $\beta_H$ is the final state velocity, $\lambda_i = \pm$ are the helicities of the 3 gluons\footnote{Here we use the same convention as~\cite{Glover-Baur}, namely that all momenta are outgoing.}, and $f_i$ indicates the different fermion species running in the loop. For simplicity, when looking at the limiting cases, we will focus on one helicity configuration, $\mathcal M_{+++}$. We will also consider only one fermion species (mass $m$, Yukawa coupling $y = \frac{m}{v}\kappa$) running around the loop and take the center-of-mass rapidity ($y^*$) to be zero.

\begin{itemize}
\item In the high-$p_T$ limit $p_T \gg m, m_H$, $\mathcal M_{+++}$ contains single- and double-logarithms of the form~\cite{crazy,Glover-Baur}
\be
\mathcal M_{+++}\Big|_{p_T \gg m,m_H} \propto \frac{m^2\,\kappa}{p_T} \Big( A_0 + A_1\, \ln\Big(\frac{p^2_T}{m^2}\Big) + A_2\, \ln^2\Big(\frac{p^2_T}{m^2} \Big) \Big),
\label{eq:mppp}
\ee
where $A_0, A_1, A_2$ are combinations of constants and $m$-independent logarithms such as $\ln{\Big(\frac{p_T}{m_H} \Big)}$\footnote{Relaxing the assumption of $y^* = 0$, these coefficients will depend on the center-of-mass rapidity as well.}. The $A_{0,1,2}$ are complex, since the internal fermions can go on-shell if the momenta entering the loop are sufficiently large.  In the high-$p_T$ limit, the matrix element $\mathcal M_{+++}$  clearly depends on both the mass and the Higgs coupling of the fermion in the loop. Note that $\mathcal M_{+++}$ has positive mass dimension since we have pulled out the factor of $v$ from the Yukawa coupling into the constant in front of Eq.~(\ref{eq:bigM}).
%\bea
%\frac{d \sigma}{d p_T} \propto \left| C  +  \left(\frac{y_t v}{m_h}\right)^2 \ln^2\left( \frac{\sqrt{2 \alpha } p_T}{m_t}\right) +  \left(\frac{y_T v}{m_h}\right)^2 \ln^2\left( \frac{\sqrt{2 \alpha} p_T}{M_T}\right) \right|^2 \label{logs}
%\eea
%where $C$ is contains all terms which are constants, single logs, or
%double logs not involving the top and top-partners. Here $t$ and $T$
%are the top and the top-partner respectively, $\alpha=1-\sinh^2 y$,
%$\hat s\simeq 2\alpha p_T^2$ and $\hat t=-\hat s(1+ \cos \theta)/2$. This expression obviously depends on the mass and couplings of the top-partner, a feature we are going to exploit thoroughly.  Note that in the large $p_T$ limit, $y$ is close to zero and $\hat t=\hat u=-\hat s/2$. 

%Note that there are also single-log contributions to Eq.~(\ref{logs}). For example, the process $q \bar q \to h + j$ does not contain double-logarithms, as the leading pieces are single-logs.

\item For low $p_T$, there is no dependence on the fermion mass since we are back in the $gg \to h$ limit of Sec.~\ref{sec:LET}. Instead:
\be
\mathcal M_{+++}\Big|_{m \gg p_T} \propto \kappa\, p_T
\ee
\end{itemize}

Having shown the two kinematic limits, let's  now consider the form of $\mathcal M_{+++}$ when there are two contributions, one from a lighter (EW-scale) fermion (i.e. the top quark, with mass $m_t$, coupling $\kappa_{tt}$) and one from a heavier, $\tev$-scale fermion (the top-partner, mass $M_T$, coupling $\kappa_{TT}$). When the final state has low-$p_T$, the Higgs is approximately at rest, and the the low-energy theorem applies. Raising the $p_T$, we enter an intermediate regime where the $p_T \gtrsim O(m_t)$ but $p_T \ll M_T$. Approximating the top and top-partner contributions with the high-$p_T$ and low-$p_T$ limits, respectively, the matrix element in this regime is (schematically, and up to higher order corrections):
\be
\mathcal M_{+++}\Big|_{m_t \ll p_T \ll M_T} \propto \frac{m_t^2\,\kappa_{t}}{p_T} \Big( A_{t,0} + A_{t,1}\, \ln\Big(\frac{p^2_T}{m_t^2}\Big) + A_{t,2}\, \ln^2\Big(\frac{p^2_T}{m_t^2} \Big) \Big) + \kappa_{T}\, p_T.
\label{eq:both1}
\ee
We see that the top-partner leads to a term in the amplitude proportional to $p_T$. This linear term will lead to a slower dropoff in the cross section as we push to higher $p_T$. The matrix element in this kinematic region is sensitive to the top mass and Yukawa, and the top-partner Yukawa.  There is no dependence on the top-partner mass until we go to an even higher $p_T$ regime, $p_T \gg m_t, m_H, M_T$. There,
\begin{align}
\mathcal M_{+++}\Big|_{m_t \ll p_T \ll M_T} \propto & \frac{m_t^2\,\kappa_{t}}{p_T} \Big( A_{t,0} + A_{t,1}\, \ln\Big(\frac{p^2_T}{m_t^2}\Big) + A_{t,2}\, \ln^2\Big(\frac{p^2_T}{m_t^2} \Big) \Big) \nonumber \\
& +  \frac{M_T^2\,\kappa_{T}}{p_T} \Big( A_{T,0} + A_{T,1}\, \ln\Big(\frac{p^2_T}{M_T^2}\Big) + A_{T,2}\, \ln^2\Big(\frac{p^2_T}{M_T^2} \Big) \Big).
\label{eq:both2}
\end{align}

\subsection{Matrix Element level}
\label{sec:melevel}

We now turn to numerics to study how the matrix elements change in a top-partner setup as the final state $p_T$ is increased. Since $gg$ is the dominant contribution to the total cross section, let us continue to focus on $gg \to h +g$. A useful variable is the ratio of partonic matrix elements squared:
\be
\frac{\Big| \sum\limits_{\lambda_i = \pm} \mathcal M_{t+T} \Big|^2}{\Big| \sum\limits_{\lambda_i = \pm} \mathcal M_{SM}\Big|^2} =  \frac{\Big| \sum\limits_{\lambda_i = \pm}  \Big( M_{\lambda_1\lambda_2\lambda_3}(\hat s, \hat t, \hat u, m_t, \kappa_{t}) + M_{\lambda_1\lambda_2\lambda_3}(\hat s, \hat t, \hat u, M_T, \kappa_{T}) \Big) \Big|^2 }{\Big| \sum\limits_{\lambda_i = \pm} M_{\lambda_1\lambda_2\lambda_3}(\hat s, \hat t, \hat u, m_t, 1) \Big|^2}\,.
\label{eq:mratio}
\ee

\begin{figure*}[h!]
\centering
\includegraphics[width=0.45\textwidth]{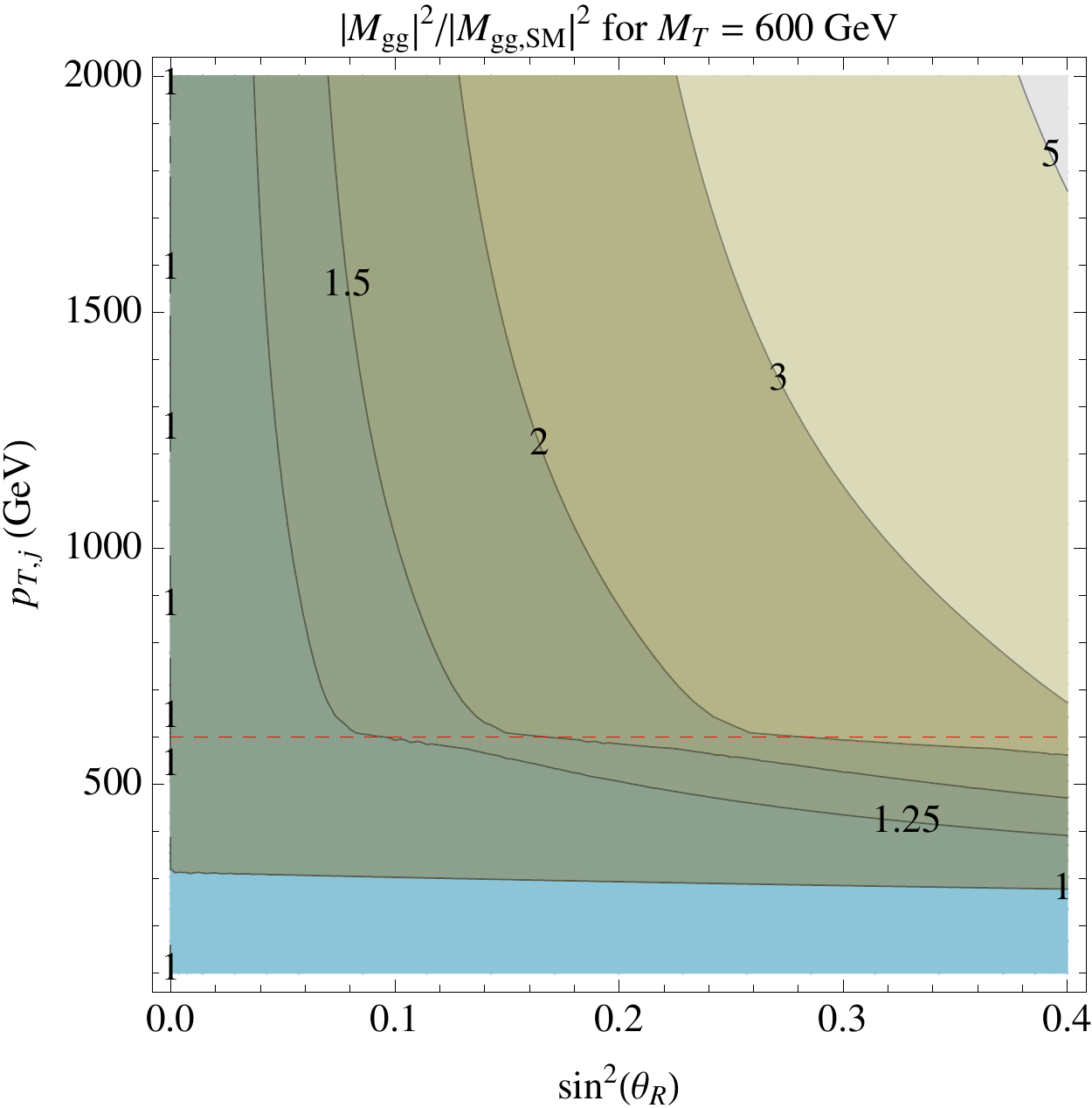}
\includegraphics[width=0.45\textwidth]{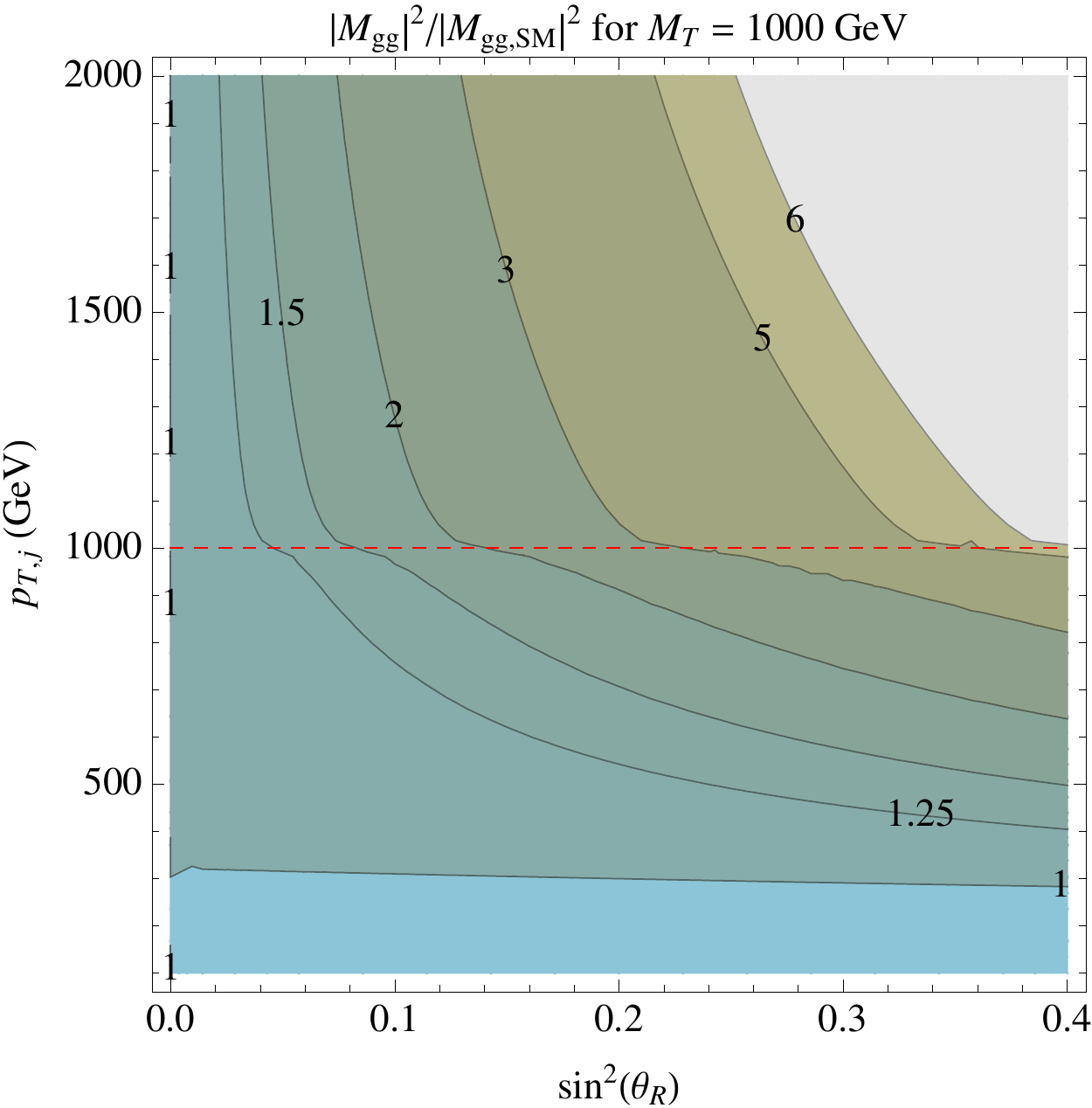} \\
\includegraphics[width=0.45\textwidth]{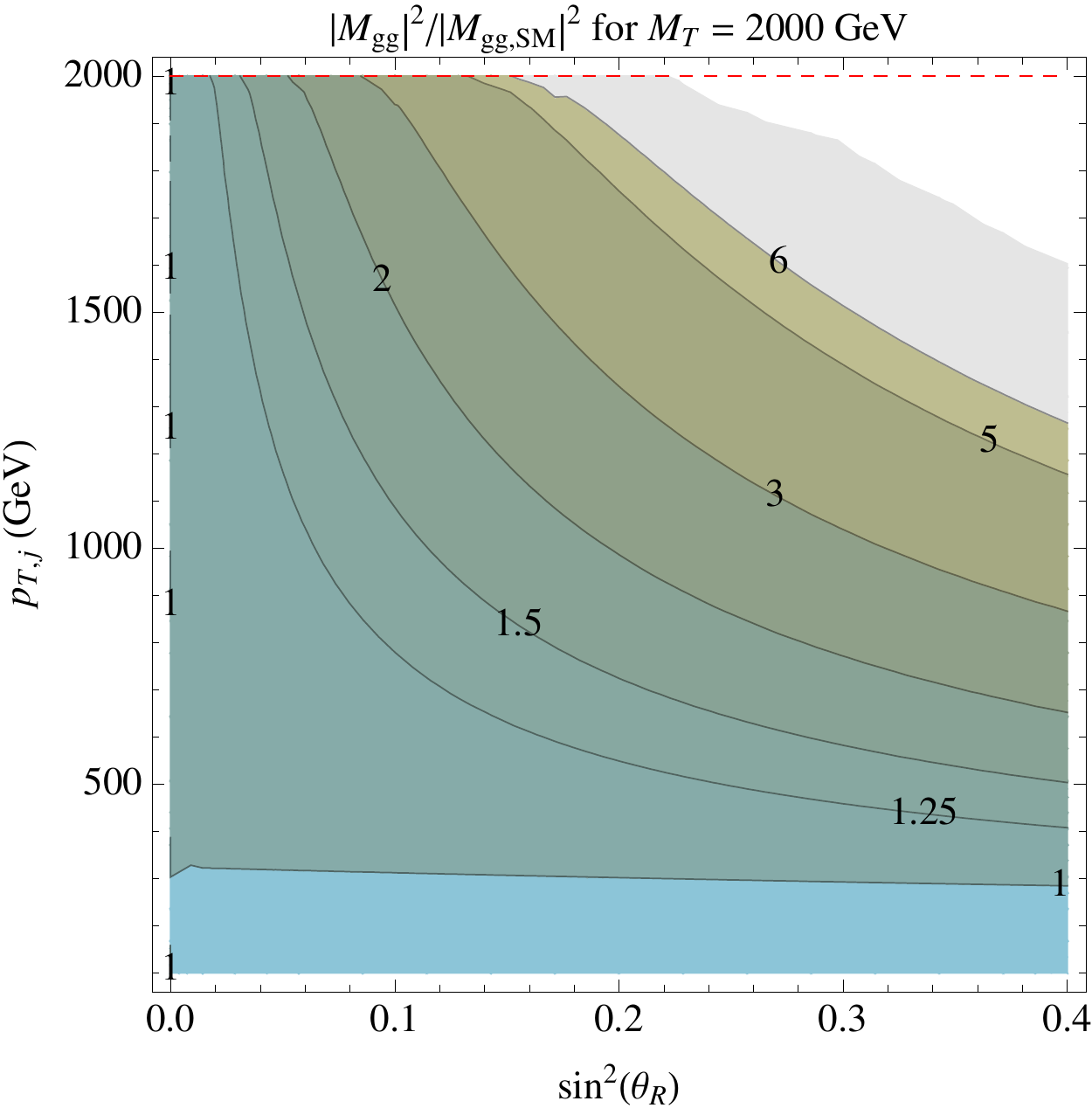}
\caption{\it Ratio of  partonic $gg \to h +g$ matrix elements squared in a theory with a 600 GeV (top left), 1 TeV (top right) and  2 TeV (bottom)  top-partner, compared to the SM value. The ratio is a function of top-mixing angle, and the $p_T$ and $y^*$ of the final state. Projecting onto $y^* = 0$ (the minimum $\sqrt{\hat s}$ for a given $p_T$), the ratio is a function of the mixing angle and $p_T$ alone. The matrix elements include all gluon polarizations. The dashed red line indicates where $p_{T,j} = M_T$.}
\label{fig:ggme_sq}
\end{figure*}

The Mandelstam variables depend on $m_H$, the $p_T$ of the Higgs (or the recoiling jet) and the rapidity of the center-of-mass frame, $y^*$. For a given $p_T$, the minimum $\hat s$ occurs when $y^* = 0$. As $\hat s = 2\,p_T\,( \sqrt{p^2_T + m^2_H} + p_T) + m^2_H$,  $\hat t = \hat u = (m^2_H - \hat s)/2$, in this kinematic region Eq.~(\ref{eq:mratio}) is a function of $p_T$, the heavy fermion mass $M_T$, and the mixing angle $\theta_R$. Fixing $M_T$  to three different values, the ratio of partonic matrix-elements squared is shown in Fig.~\ref{fig:ggme_sq} as a function of $\sin^2(\theta_R)$ and $p_T$. The shapes of the contours in Fig.~\ref{fig:ggme_sq} can be understood by the different functional forms of Eq. (\ref{eq:both1}) and Eq.~(\ref{eq:both2}): for $p_T \lesssim M_T$ (below the red dashed line) the ratios have a similar shape for all three $M_T$ values, while for $p_T \gtrsim M_T$ the contours change shape and their values depend on the $M_T$ assumed. Large ratios $\sim O(5)$ are possible, however the largest differences come at high-$p_T$ where the cross section is smallest. To gauge the effect on the full cross section we need to fold in parton distribution functions.

\subsection{Including the effect of PDFs and running}\label{inclpdf}

We now move onto the effect of including scale and PDF effects. This has been done by adapting Herwig~\cite{herwig} amplitudes to include contributions from a top-partner. The modified matrix elements were then interfaced with HOPPET \cite{hoppet} and LHAPDF \cite{lhapdf} to generate the distributions. We also implemented the top-partner in MCFM~\cite{MCFM} to check our results\footnote{Finite-mass effects in loops are implemented also in Pythia~\cite{pythia}, POWHEG~\cite{powheg} or MC@NLO~\cite{mcatnlo}, so we could have used any of those programs instead of Herwig.}. For the SM, our calculation includes the effects of both the bottom and top quarks; for the top-partner scenarios we include the top, top-partner (with $\theta_R$ dependent Yukawa couplings), and bottom quark contributions. The  differential $p_T$ distribution is shown below in Fig.~\ref{mcfmcomp} for the SM and six top-partner scenarios --  three different $M_T$ values and two different $\sin^2(\theta_R)$ values. 
\begin{figure}[h!]
\centering
\includegraphics[width=0.49\textwidth]{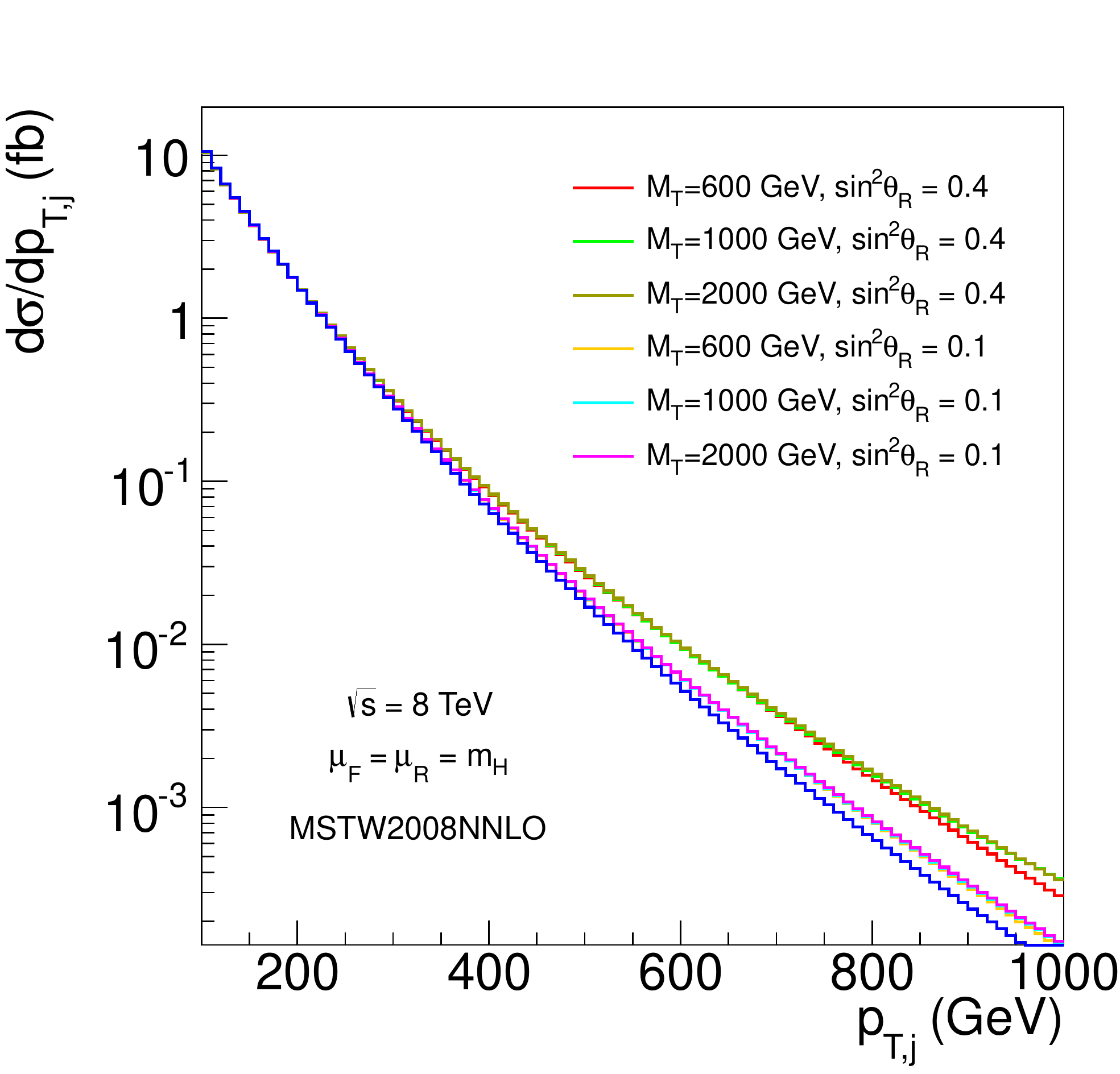}
\includegraphics[width=0.49\textwidth]{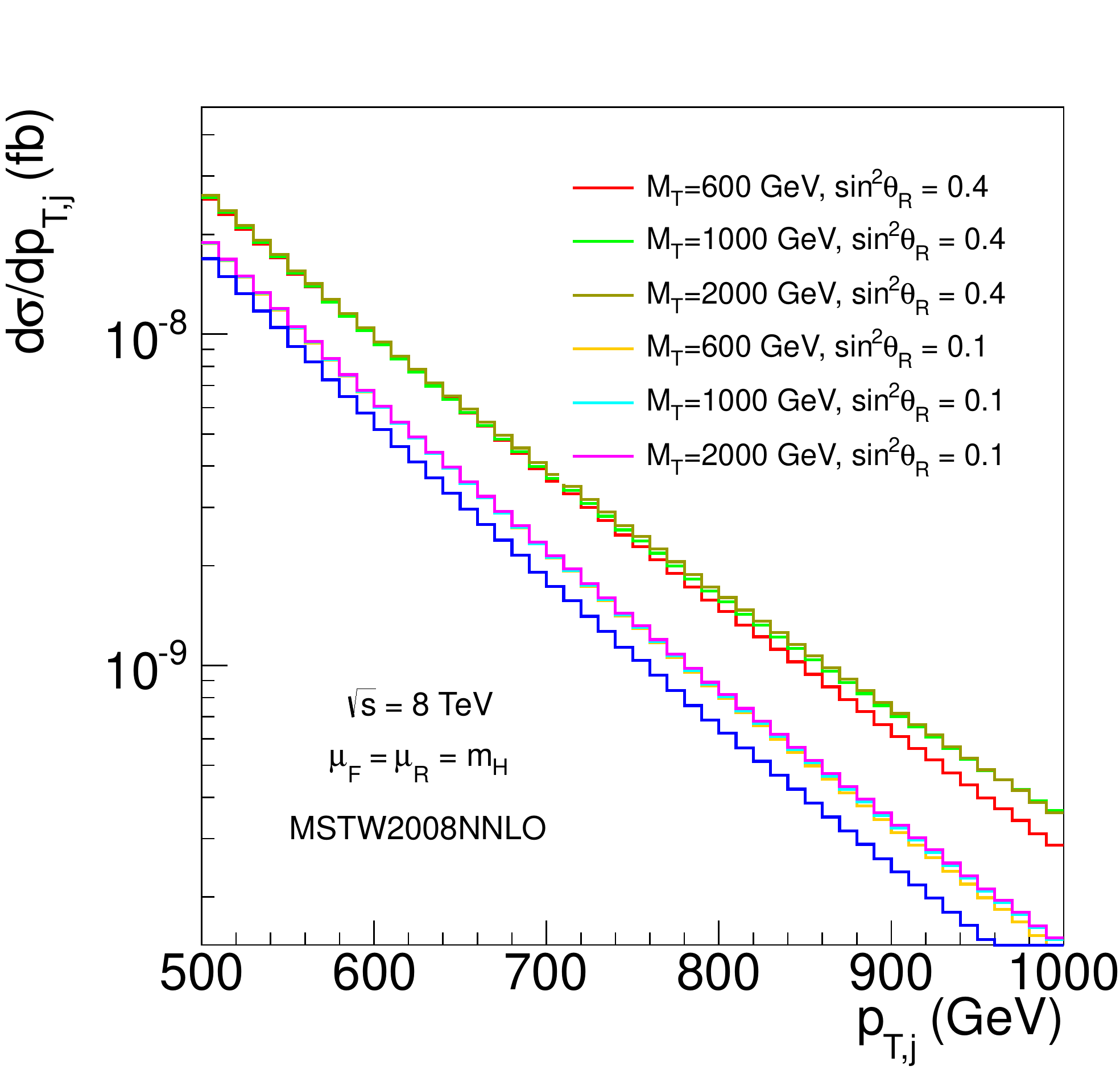}
\caption{(Left panel) differential cross section $d\sigma/dp_T$ at a $\sqrt s  = 8\, \tev$ LHC for the SM (top and bottom quarks) in blue, and including a top-partner. Three different top-partner masses are shown, $600\, \gev. 1\, \tev$ and $2\, \tev$ and two different top-mixing angles $\sin^2(\theta_R) = 0.1, 0.4$. (Right panel) same spectra, zoomed in to the high-$p_T$ range $500\, \gev - 1\, \tev$.}
\label{mcfmcomp}
\end{figure} 
This plot exhibits the same features we saw at the partonic level, though diluted by the PDFs. First, as dictated by the low-energy theorem, all top-partner scenarios converge to the SM result at low-$p_T$. Second, as suggested by the analytic results in Sec.~\ref{sec:general}, the $p_T$-spectra in top-partner scenarios is harder than the SM. Finally, the spectra for a given mixing angle are not sensitive to the top-partner mass until the final state $p_T \sim M_T$.

\begin{figure}[h!]
\centering
\includegraphics[height=7cm]{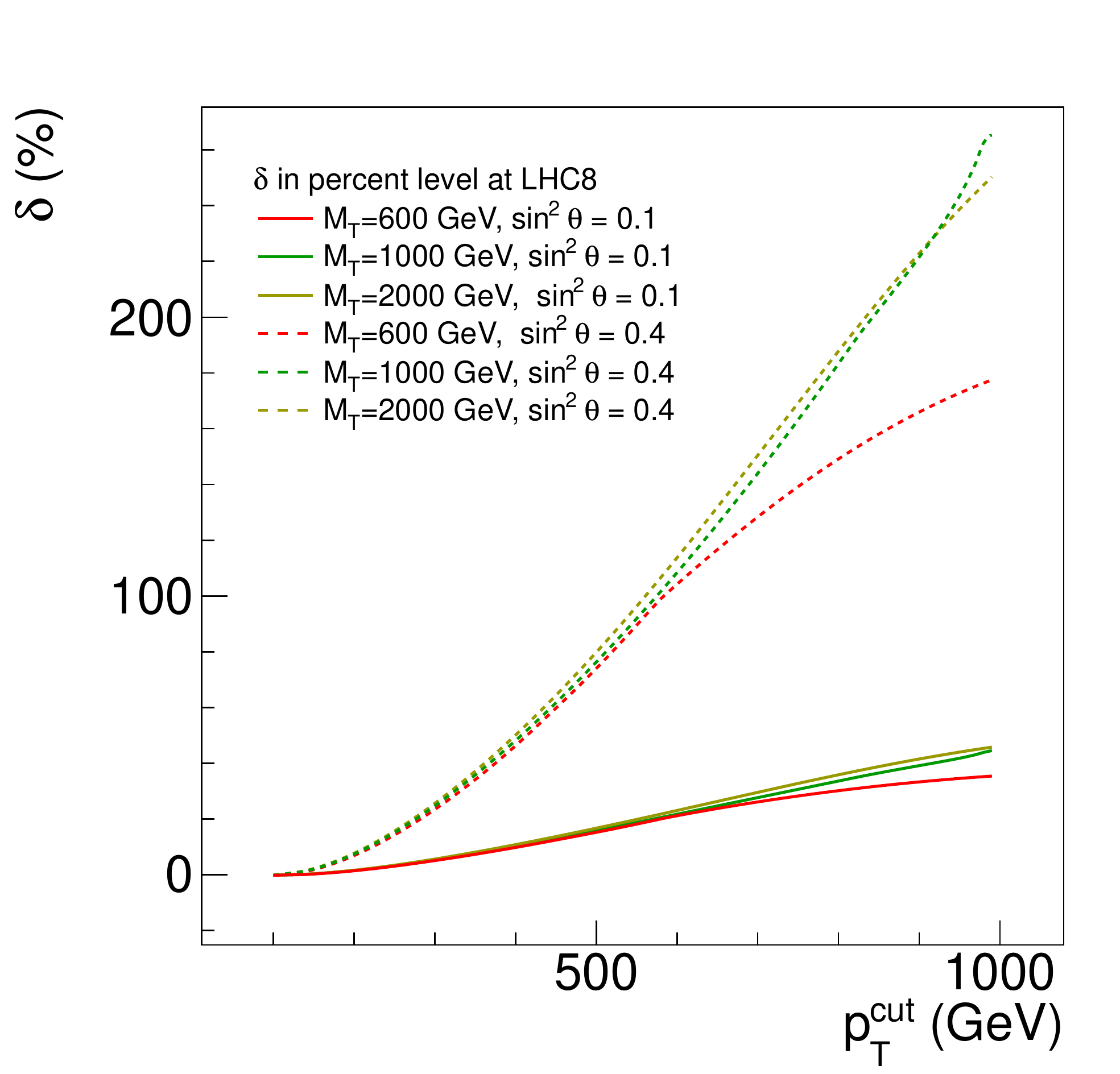}
\includegraphics[height=7.cm]{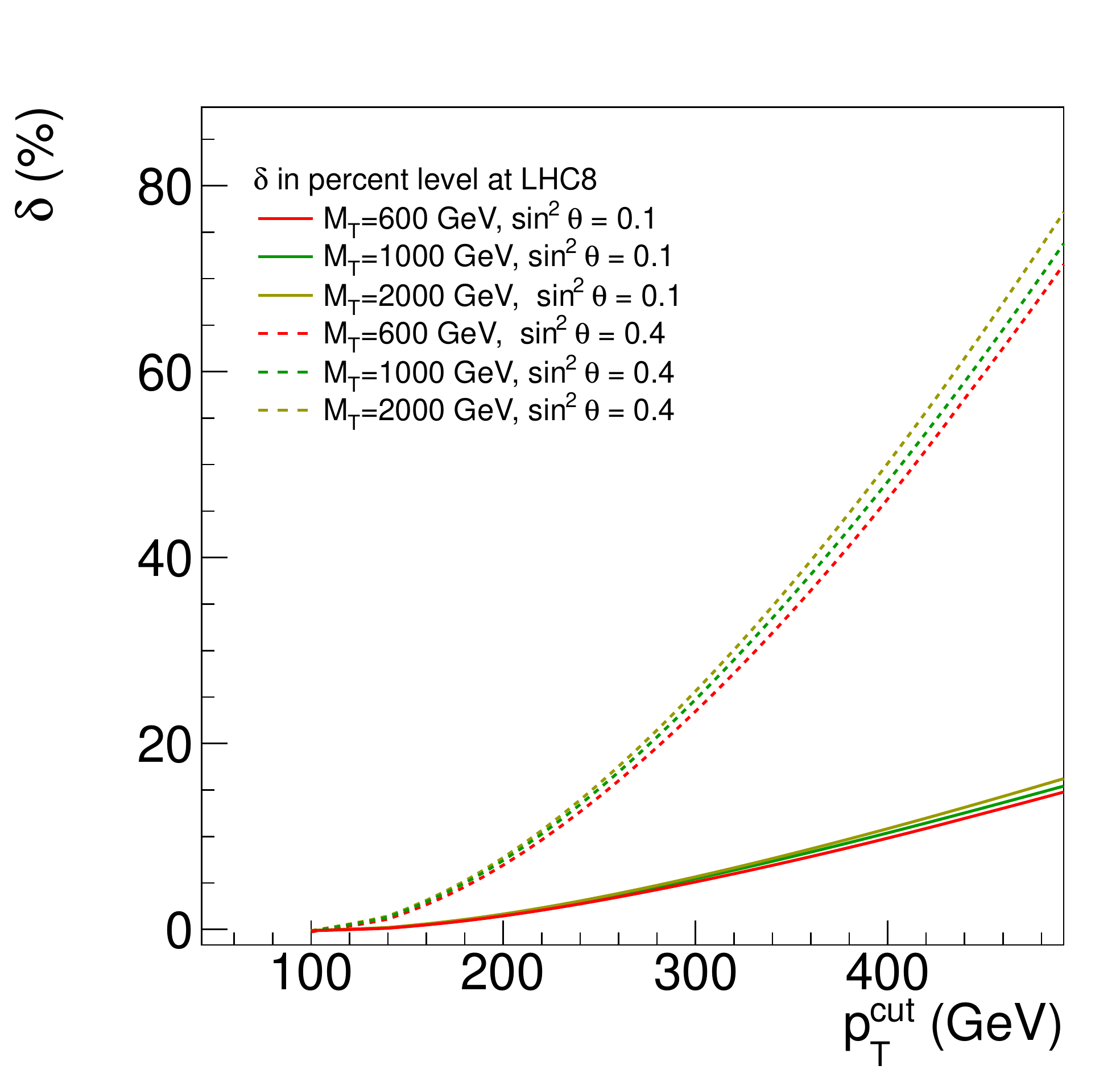} 
\caption{\it (Left panel) $\delta$ as a function of $p_T^{\rm cut}$ for different values of $M_T$ and the mixing angle. (Right panel) A zoom in the interesting range of  $p_T$.  These plots have been generated with $\mu=\frac{1}{2} (p_T+\sqrt{p_T^2+m_h^2})$ and a $\sqrt s = 8\, \tev$ LHC.}
\label{deltaLHC8}
\end{figure}

The difference in the $p_T$ spectrum between the SM and a theory with a top-partner is our main result. The full $p_T$ spectrum is, however, an experimentally difficult quantity to measure since the higher $p_T$ bins will suffer from low statistics. A similar, though perhaps experimentally more tractable observable is the net Higgs plus jet cross section for all events that satisfy a given $p_T$ cut, i.e.
\bea
\sigma (p_T > p_T^{\rm cut}) = \int_{p_T^{\rm cut}} d p_T \frac{d \sigma}{ d p_T} \ .
\label{spt}
\eea 
Using $\sigma(p_T > p_T^{\rm cut})$, we define a new variable $\delta$,
\bea
\delta (p_T^{\rm cut}, M_T, \sin  \theta) = \frac{\sigma_{t+T} (p_T > p_T^{\rm cut}, \mu, M_T, \sin  \theta)-\sigma_{t} (p_T > p_T^{\rm cut}, \mu)}{\sigma_t (p_T > p_T^{\rm cut}, \mu)} \ .
\eea
which encapsulates the effect of a top-partner in the cross section. Here,  $\sigma_{t+T}$ is the cross-section in a theory with a top-partner of a given mass and mixing angle, while $\sigma_t$ is the cross-section for the SM.  In Fig.~\ref{deltaLHC8} we show the value of $\delta$ as a function of $p_T^{\rm cut}$ for different values of $M_T$ and the mixing angle. Obviously, the effect increases with the top-mixing angle. As in the differential distributions, heavier top-partners lead to a harder $p_T$ spectrum, but the effect $\delta$ is negligible until $p_T > M_T$. To generate this plot, we have taken $\mu_R=\mu_F=\mu=\frac{1}{2} (p_T+\sqrt{p_T^2+m_h^2})$ and $\sqrt s = 8\,\tev$. 

\begin{figure}[h!]
\centering
\includegraphics[height=7.cm]{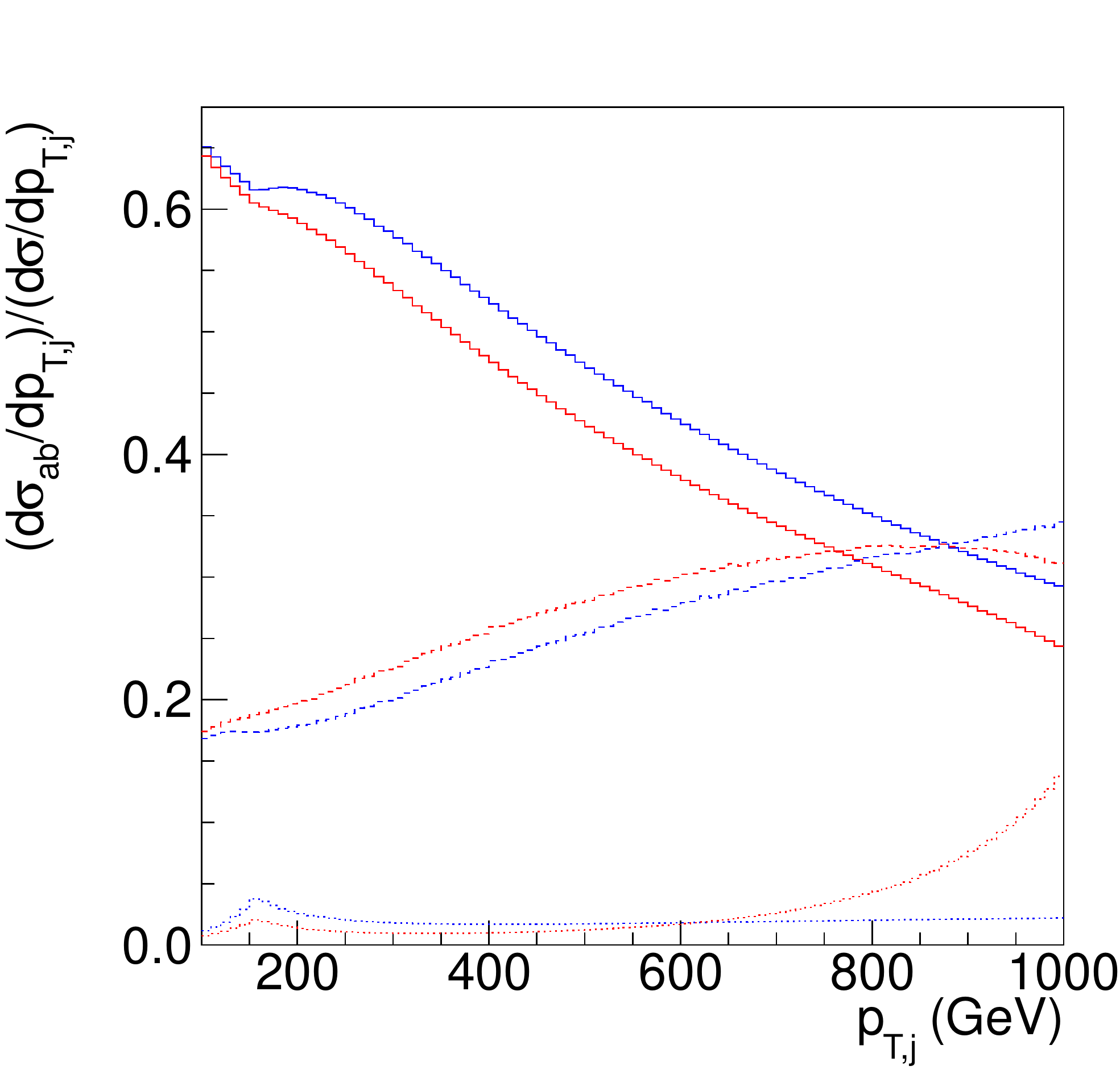}
\caption{\it Breakdown of the differential cross section $d\sigma/d p_T$ into different initial state channels, $d\sigma_{i}/d p_T$, where $i = $  $gg$ (solid), $qg + \bar q g$ (dashed) and $q\bar q$ (dotted). The blue (red) lines correspond to the SM (top-partner) theory. The top-partner in this plot corresponds to $M_T = 1\, \tev$ and $sin^2({\theta}_R)=0.4$. The contribution from $gq + g\bar q$ is not shown (thus the sum does not equal 1.0) since it is identical to $qg + \bar q g$. }
\label{breakd}
\end{figure}

While gluon-initiated subprocess dominate $pp \to h+j$ for low $p_T$, it is interesting to see how the breakdown of the cross section into partonic subprocesses changes as we increase the $p_T$.  In Fig.~\ref{breakd} we plot the ratio
\bea
 \frac{d\sigma_{i}}{d p_T}/\frac{d\sigma_{\rm tot}}{d p_T},\quad  \, i =  gg, gq + \bar q g, {\rm or} q\bar q 
 \eea
in the SM and in the theory with a $1\, \tev$ top-partner (here, $\frac{d\sigma_{\rm tot}}{d p_T}$ is the differential distribution including all channels in the respective theory).
The dominant cross section corresponds to $gg$ for jet $p_T\lesssim$ 800 GeV, after which $qg$ becomes the dominant subprocess. The crossover is delayed in the case of a theory with a top-partner respect to the SM, as the former exhibits a harder spectrum.
Note also that the $qg$ and $q\bar q$ initial states do depend on the quark mass. For example, in the right-hand diagram in Fig.~\ref{fig:feyn}, the dependence on the quark in the loop can be understood as the $t$-channel gluon virtuality enhancing the double-logarithmic structure in the matrix element. The sharp features in the $\bar q q$ subprocess at  $p_T \sim m_T$ and $p_T \sim M_T$ come from a resonant enhancement in the loop functions near $\hat s \sim 4\,p^2_T \sim 4m^2_t$ or $\hat s \sim 4\,M^2_T$ respectively. Had we plotted to $p_T > 1\, \tev$, the $\bar q q$ fraction in the top-partner scenario would shrink again.\\

Although we have calculated loops of top quarks and top-partners, our $pp \to h+j$ calculation is still a lowest-order calculation. Being a lowest order (LO) result --  especially given that the cross section depends on $\alpha^3_s$ -- one immediate worry is that our result may be highly dependent on the scale choice and the choice of PDF.  However, provided we look at a ratio of cross sections, such as $\delta(p^{\rm cut}_T)$, one might expect most dependence on these input choices should drop out. We have confirmed this intuition with cross-checks. First, calculating $\delta(p^{\rm cut}_T)$ for three different values of the factorization and renormalization scheme, $\mu_R = \mu_F =  \mu=\left(p_T+\sqrt{p_T^2+m_h^2}\right)/2$, $\sqrt{p_T^2+m_h^2}$ and $m_H$, we find the difference in the ratio between the three schemes, i.e. $\delta(p^{\rm cut}_T, \mu)/\delta(p^{\rm cut}_T,\mu')$ is below the percent level. Next, we verified the stability of $\delta(p^{\rm cut}_T)$ under changes in the PDF schemes by comparing $\delta(p^{\rm cut}_T)$ calculated with two different PDF sets. Using top-partner parameters $M_T = 1\, \tev, \sin^2(\theta_R) = 0.4$, the ratio of $\delta(p^{\rm cut}_T)$ calculated with {\tt MSTW2008nlo68cl} PDFs~\cite{Martin:2009iq}  to $\delta(p^{\rm cut}_T)$ calculated using {\tt cteq6mE}~\cite{Nadolsky:2008zw} is shown below in Fig.~\ref{pdfcomp}. The effect is less than 2\% in the range of $p_T$ we will consider.

\begin{figure}[h!]
\centering
\includegraphics[width=0.5\textwidth]{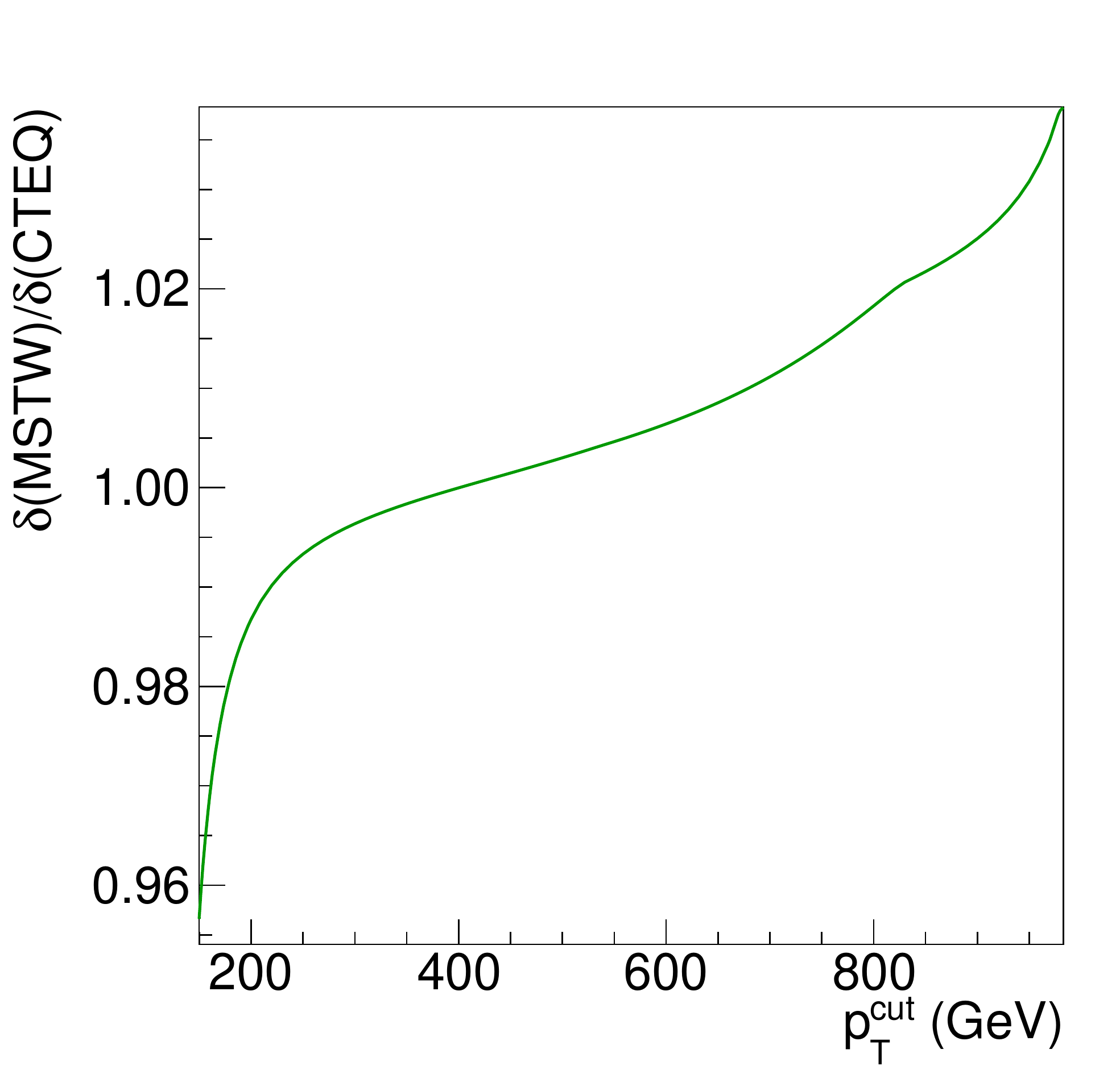} 
\caption{\it The ratio of  $\delta(p^{\rm cut}_T)$ calculated with  {\tt MSTW2008nlo68cl} parton distribution functions to $\delta(p_T)$  calculated with {\tt cteq6mE}. The top-partner used for calculating $\delta(p_T)$ has mass $1\, \tev$ and mixing angle $\sin^2(\theta_R) = 0.4$. All distributions were generated using $8\,\tev$ LHC parameters. }
\label{pdfcomp}
\end{figure}

\begin{figure}[h!]
\centering
\includegraphics[width=0.5\textwidth]{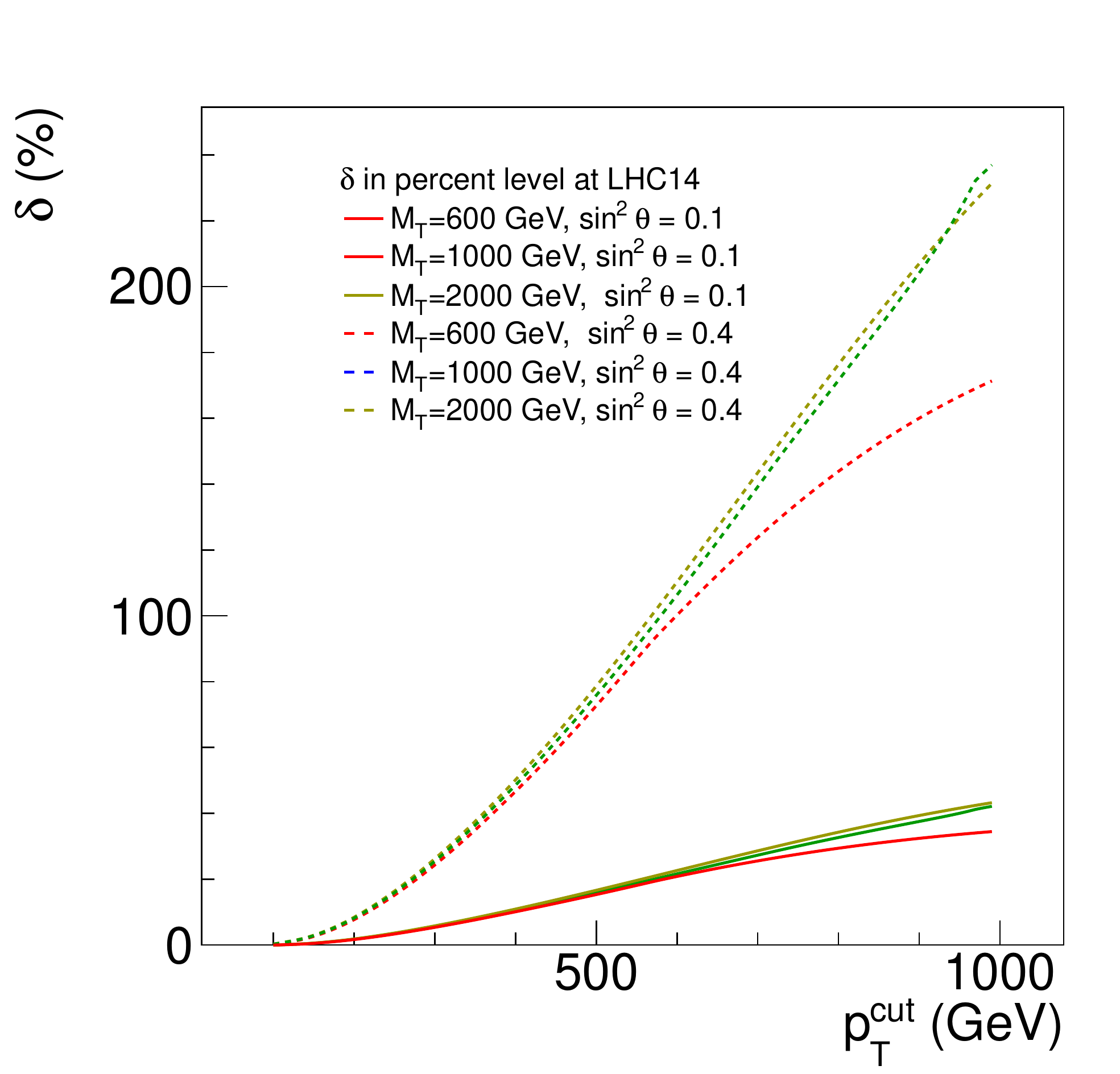}
\caption{\it $\delta$ as a function of $p_T^{\rm cut}$ for different values of $M_T$ and the mixing angle for $\sqrt s = 14\, \tev$.}
\label{del14}
\end{figure}

We move on to study the effect of the collider energy by comparing the results for $\sqrt s = 8\,\tev$ and $\sqrt s = 14\,\tev$. The quantity $\delta(p^{cut}_T)$ is shown in Fig.~\ref{del14}. Comparing with the same quantity at $\sqrt s = 8\, \tev$ Fig.~\ref{deltaLHC8}, one can see that the ratio does not depend strongly on the energy of the collider.

Finally, a comment on the dependence of the result on the rapidity acceptance for the jet. The topology we are looking at, with a Higgs recoiling against a high-$p_T$ jet tends to produce very central events. This is just because at high $p_T$ there is not enough phase space to produce high rapidity jets. Indeed, our Herwig implementation, in which we have integrated over all rapidities, is in agreement with MCFM with a cut $|\eta|<5$. We have checked in MCFM that moving the cut on jet rapidity from $|\eta|<5$ to $|\eta|< 2.5$, which corresponds to the acceptance of the CMS and ATLAS central trackers, does not alter our results.  

%%%%%%%%%%%%%%%%%

\section{Stability against higher order corrections and experimental uncertainties}
\label{HOC}

In Sec.~\ref{Hj} we discussed the stability of the results when changing the renormalization scale and PDF sets, finding that the effect is at the percent level. 
In this section we will focus on the effect of adding higher order corrections and experimental uncertainties.

Currently, there is no available computation of Higgs plus jet at next-to-leading order (NLO) including finite mass effects\footnote{In fact, there exists a  calculation of Higgs plus one jet at NLO, but contains only top-mass effects in the heavy-top limit up to $1/m_t$ corrections, so it can be used only at moderately low $p_T$~\cite{Harlander:2012hf}.}. This calculation is beyond the scope of this paper, but given its importance for constraining new physics, one would hope that it becomes available in the near future.  Given this situation, the best one can do is to evaluate the NLO effects, differentially, in the infinite top mass limit. We have evaluated the K-factor, the LO and NLO Higgs plus jets using MCFM~\cite{MCFM} in the infinite top mass limit in the differential distribution $d\sigma/d p_T$. The K-factor is rather flat (roughly $O(2)$) as a function of $p_T$ for $\mu=\sqrt{p_T^2+m_H^2}$, but has a slope for $\mu=m_H$. 

We expect the higher order corrections to produce changes in shape once the finite mass effects are taken into account. Nevertheless, as our observable is an integrated cross section, dominated by the region near the $p_T$ cut, we expect  higher-order corrections to just amount to an overall K-factor, although this expectation should be corroborated with a explicit calculation. Moreover, one should aim to obtain as much information as possible from the {\it differential} cross section, whereas in this paper we have to limit ourselves to an integrated cross section with a $p_T$ cut. With a NLO calculation with finite mass effects, the differential distribution would become a more powerful tool to disentangle new physics.

The most important experimental uncertainty for our observable would be energy and momentum smearing of the Higgs or the recoil jet. As we are using integrated cross sections as in Eq.~(\ref{spt}), the effect of smearing would affect the region near the cut. Similarly, the effect of the underlying event would also produce some momentum smearing, although we expect it would be negligible at $p_T>$ 100 GeV~\cite{giulia}.  Therefore, at least for $p_T$ cuts greater than $\sim 200\,\gev$, we believe experimental effects should be small and will affect the SM and top-partner scenarios in a similar way. 

%Regarding the experimental uncertainties, the effect would mostly come from the smearing of the energy and momentum of the jet, or the Higgs.

%%%%%%%%%%%%%%%%%%%%%%
\section{Mass limits on top-partners}
\label{seclimits}

In this section we  study the sensitivity of the $14\,\tev$ LHC to the top-partner mass and coupling. The events we are focusing on are characterized by a high-$p_T$ jet plus a Higgs boson. 

Given a particular Higgs plus jet final state and some amount of luminosity, we can set limits on the top partners by comparing two hypothesis: SM Higgs plus jet production vs. Higgs plus jet production in a top partner scenario, where the latter hypothesis is a function of $M_T$ and $\sin^2({\theta_R})$. For simplicity, and since there is no dedicated CMS/ATLAS search in Higgs plus hard jet to work off of, we will quantify the difference between the two hypothesis with the variable
\bea
\textrm{ Significance } = \frac{S}{\sqrt{S_0}}\,,
\eea
where $S$ is the signal
\bea
S = \left( \sigma_{t+T} (p_T> p_T^{\rm cut}) - \sigma_{t} (p_T> p_T^{\rm cut})\right) \times {\cal L} \ ,
\eea
and $S_0$ is the SM piece, $S_0=\sigma_{t} (p_T> p_T^{\rm cut}) \times {\cal L}$. We claim sensitivity to rule out a top-partner at the 95 \% confidence level if $S/\sqrt{S_0}$ at luminosity $\cal L$ is bigger than 2.0. 

This test statistics is only approximate as it assumes that the SM background can be completely removed. This is a reasonable assumption in the clean leptonic and photon final states. For the higher rate, hadronic Higgs decay modes the SM background is more problematic, though the requirement of a hard jet in the event is a useful handle for suppressing background. Dedicated studies of the backgrounds in all Higgs final states for Higgs plus hard jet events are well motivated, but beyond the scope of this paper.

Our test statistics also assumes that the cut efficiency for the SM and new physics Higgs plus jet events is the same, and that the Higgs branching ratios are not modified by new physics\footnote{As long as the top-partner is beyond threshold, the possible modification of Higgs branching ratios from the SM is a question which does not directly depend on the top-partner.}. A final caveat in our significance measure is that we use LO cross sections only. As we mentioned in Sec.~\ref{HOC}, the complete, mass-dependent higher order corrections are not known yet and may carry some non-trivial fermion mass and $p_T$ dependence.

%
% and the SM backgrounds will depend on which final state we look at, we 
%
%, but the limits would also depend on the final state to which the Higgs decays to. In any case, once the Higgs would have been identified via a mass reconstruction, the main background to our top-partner signal would be the SM Higgs plus jets production. Hence, we set a limit as

%To make this plot, we assumed that the hard jet is a handle which may allow to reconstruct the Higgs in states with higher rates (hadronic), besides the very clean leptonic and photon final states. The limit would be modified if one considered a subset of Higgs final states ($Br$), and the efficiency to identify the Higgs in those ($\epsilon$). The efficiency $\epsilon$ would be a function of the $p_T$ of the Higgs, and hence different for the SM ($\epsilon^{SM}$) and new physics ($\epsilon^{t+T}$)~\footnote{ The branching ratio of the Higgs to a specific final state could also depend on the new physics, but as long as the top-partner is beyond threshold, the possible modification of $Br$ from the SM is a question which does not directly depend on the top-partner.}. One would then re-scale the significance as
%\bea
%\frac{S}{\sqrt{B}} \to \frac{\epsilon^{t+T}\, \sqrt{Br}}{\sqrt{\epsilon^{SM}}} \, \frac{S}{\sqrt{B}}
%\eea
%

\begin{figure}[h!]
\centering
\includegraphics[height=6cm]{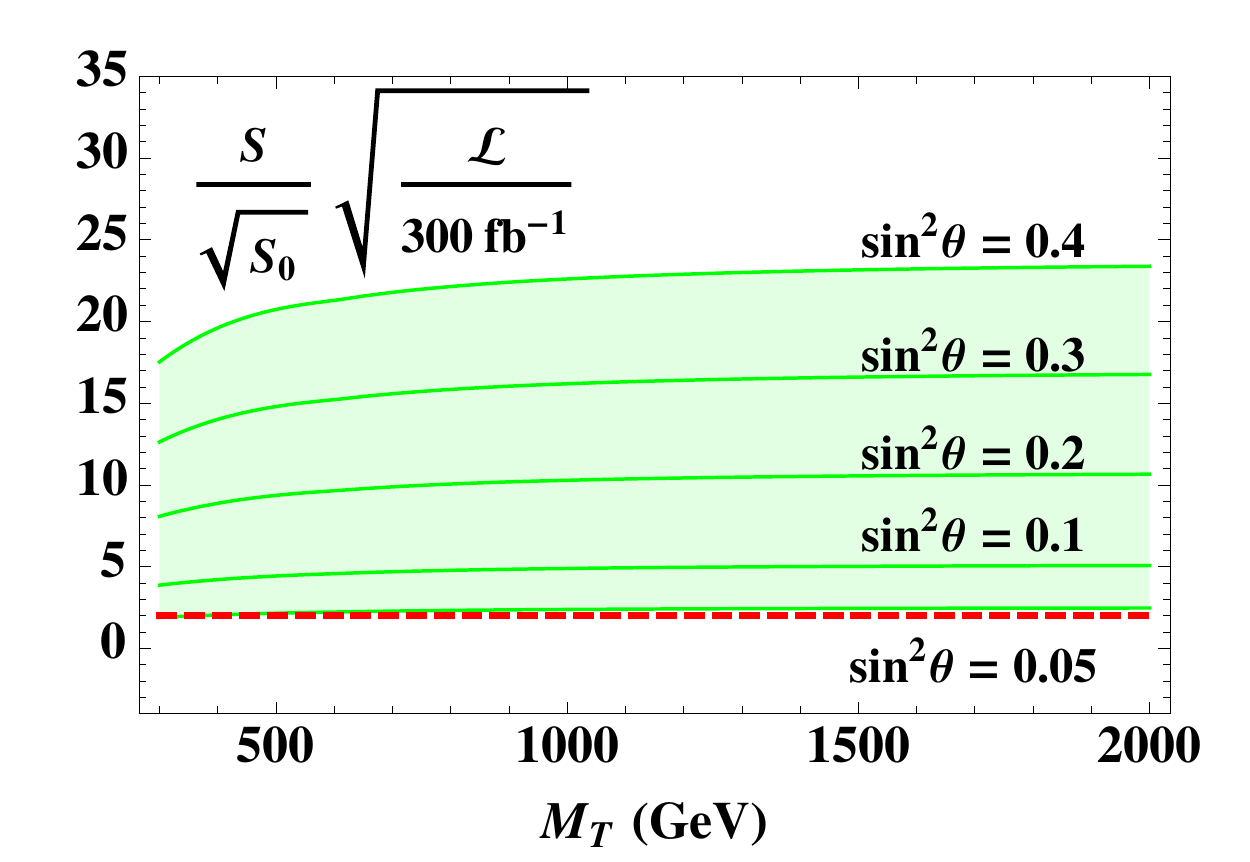}
\caption{\it Significance as a function of the mixing angle for a standard luminosity fo 300 fb$^{-1}$.}
\label{limitsm}
\end{figure}

In Fig.~\ref{limitsm}, we show the significance as a function of the mixing angle for a standard luminosity fo 300 fb$^{-1}$. With mixing angles $\sin^2 (\theta_R) \gtrsim 0.05$, one would have sensitivity in a range from around 300 GeV to above 2 TeV. 
Recall that in Fig.~\ref{fig:atf} we showed the translation between the mixing angle and the top-partner parametrizations in the literature. A limit of $\sin^2 (\theta_R)$ at 0.05, is equivalent to a limit on the scale of breaking $f$ for a fixed value of $a_T$. For example, 
\bea
\sin^2(\theta_R) < 0.05 \Rightarrow f > 1.6 \textrm{ TeV, for } \lambda_T \simeq y_t \ . 
\eea

In Sec.~\ref{inclpdf}, we showed that $\delta(p^{\rm cut}_T)$ is very stable against changes in definitions of renormalization scale and PDF sets.  We have checked that the quantity $S/\sqrt{S_0}$ is also rather stable. To do so, we define
\bea
\Delta(\omega_1,\omega_2) = \frac{S/\sqrt{S_0}(\omega_1)-S/\sqrt{S_0}(\omega_2)}{S/\sqrt{S_0}(\omega_1)+S/\sqrt{S_0}(\omega_2)},
\eea
where $\omega_i$ is a label for the choice of running parameters. The value of $\Delta$ for the same two choices of PDF schemes mentioned in Sec.~\ref{inclpdf} is shown below in Fig.~\ref{Deltafig}. As before, the effect at the sub-percent level. We have also checked agains changes in renormalization scales and PDF sets within a PDF scheme. 

\begin{figure}[h!]
\centering
\includegraphics[height=5.5cm]{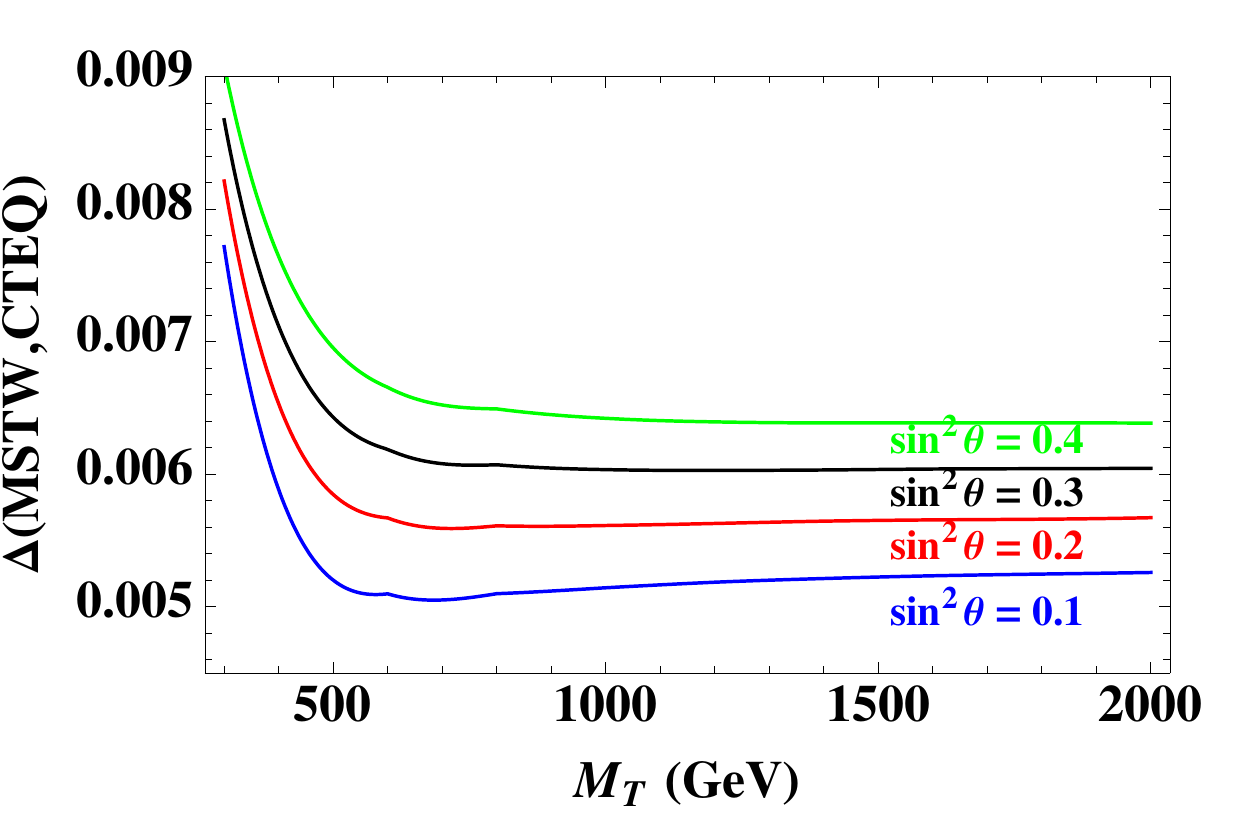}
\caption{ \it  $\Delta$ for two choices of PDF schemes with a cut on $p_T>$ 200 GeV.}
\label{Deltafig}
\end{figure}

From Fig.~\ref{limitsm}, we see that the sensitivity curves are fairly flat, indicating that the $S/\sqrt{S_0}$ is mainly sensitive to the coupling. To see the difference between higher top-partner masses, we would need to look at higher-$p_T$, where there is simply not enough rate at $\sqrt s = 14\, \tev$. This fact makes the Higgs plus jet search quite complementary to traditional $pp \to  T \bar T$ top-partner searches, where the production the rate is set by $M_T$ alone. The decay of top-partners is more model dependent. However, at least in simple setups, the decay is completely governed by "Goldstone-equivalence"  and is thus independent of the $T\bar T h$ coupling.

\begin{figure}[h!]
\centering
\includegraphics[height=5.1cm]{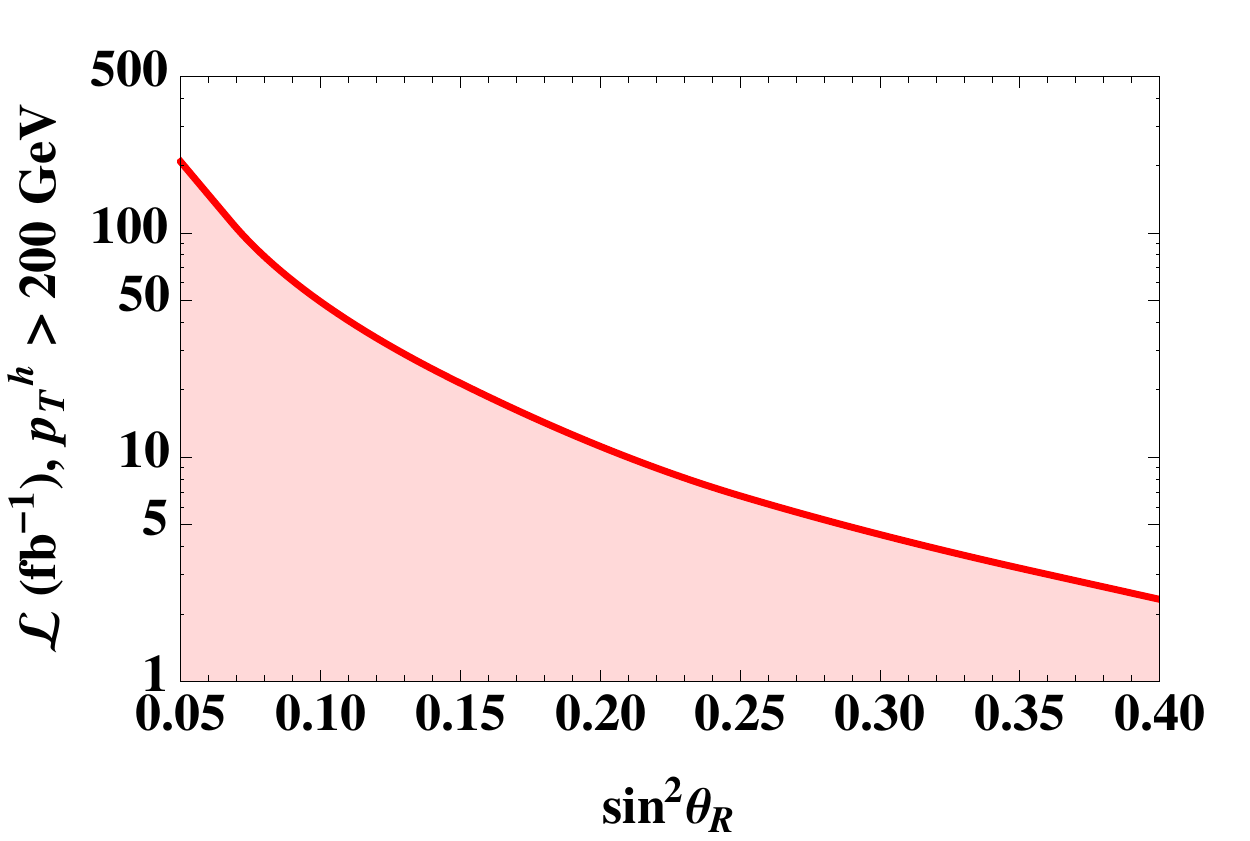} \includegraphics[height=5.1cm]{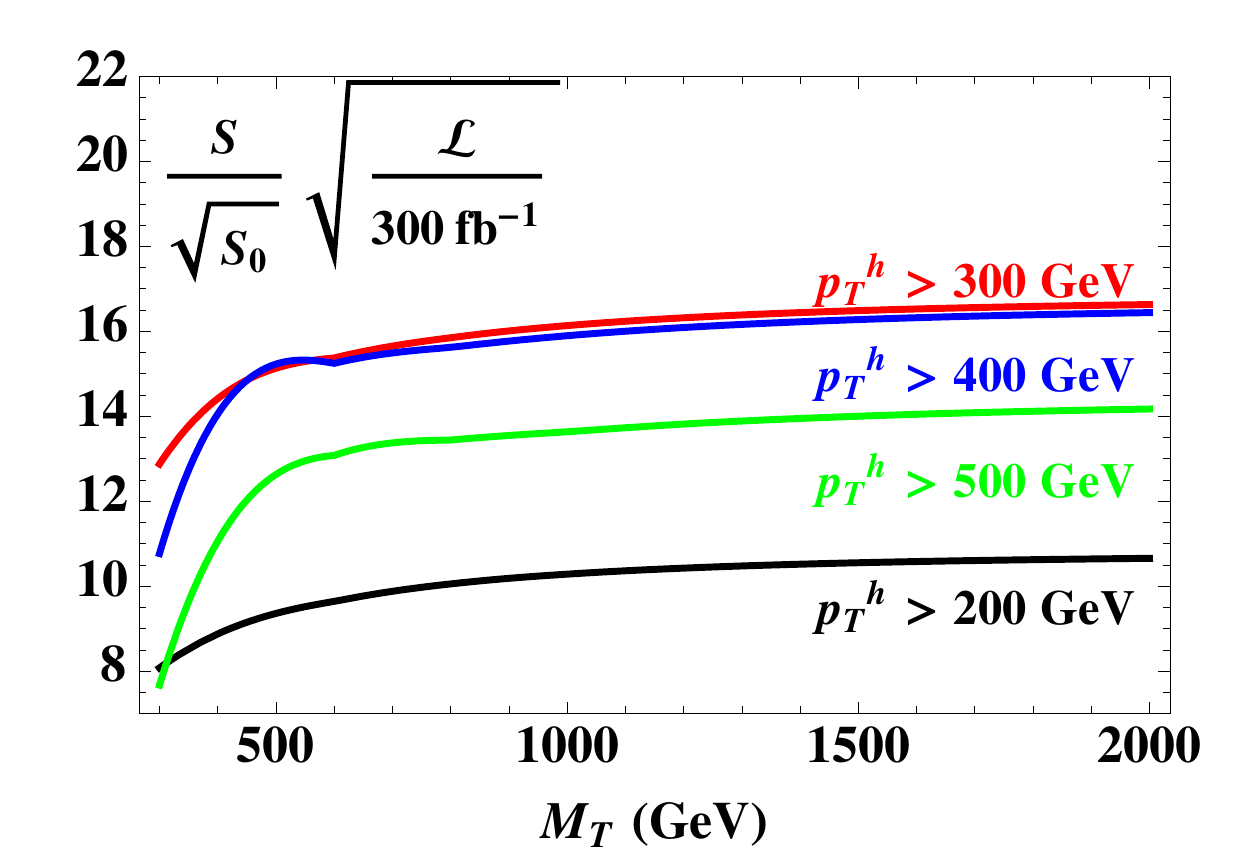}
\caption{\it (Left panel) Luminosity (in fb$^{-1}$) required to rule out  a mixing angle at the 2$\sigma$ level with a cut on $p_T>$ 200 GeV. (Right panel) Effect of raising the $p_T$ cut on the significance, for $\sin^2(\theta_R)$ =0.2.}
\label{pTvs}
\end{figure}

As the sensitivity is rather flat with $M_T$ for $M_T\simeq$ 600 GeV, one can plot the luminosity required to set a exclusion as a function of the top-mixing angle alone. This is shown in the left panel of Fig.~\ref{pTvs}, where we have chosen a cut on $p_T$ of 200 GeV. In the right panel of Fig.~\ref{pTvs} we show the effect of changing this cut for $\sin^2(\theta_R) = 0.2$. As the cut increases, the sensitivity does increases until at about $p_T\simeq $ 400 GeV, the cut is too hard and the sensitivity starts decreasing. 

\section{Conclusions}

In this paper we have presented a first step to search for top-partners in events where the Higgs is produced in association with hard jets. This topology avoids the well-known low-energy cancellation acting the $hgg$ coupling when the Higgs is a pseudo-Goldstone boson that renders the $g g \to h$ process  insensitive to the mass and coupling of the top-partner. Our analysis is motivated by these type of models, but it just relies on the presence of a top-partner with couplings to the Higgs coming from electroweak symmetry breaking\footnote{For instance, our result applies to extra dimensional models such as Ref.~\cite{Cheng:1999bg,Rius:2001dd,Carena:2006bn,Contino:2006qr,Burdman:2007sx}, and some topcolor models~\cite{Hill:1991at}} .

We have worked out the dependence of the spectrum on the top-partners using variables with are not directly the differential distribution, but integrated distributions with a cut on $p_T$.
We checked that the results at leading order are stable against
choices of renormalization scales and PDF sets. We discussed what
would be the effect of including NLO corrections. Unfortunately, no
NLO computation is available in the finite mass limit. We did check
that in the infinite mass limit the K-factor on the differential
distribution is flat for appropriate choices of the renormalization
scale. 

Finally, we have assigned a significance for a top-partner signal, finding that 200 fb$^{-1}$ of data may access very low mixing angles $\sin^2(\theta_R)<$0.05, for a large range of top-partner masses. These results are certainly encouraging and warrant more dedicated study. 

Furthermore, more information could be obtained by looking at the differential distribution, as opposed to the integrated one. This study would require an excellent understanding of the NLO corrections of this distribution, a calculation we hope will become available in the near future.

\section*{Acknowledgements}

The work of AB and VS is supported by the Science Technology and Facilities Council (STFC) under grant number ST/J000477/1. 

\appendix
\section{Choices in Model Space}\label{choices}

Although our study would be model independent, one can map the parameter space of Composite and Little Higgs models into our setup. In particular, we show examples  in the minimal coset $SO(5)/SO(4)$~\cite{CHMs} and study top-partners in the singlet and fundamental representation of $SO(4)$, which we denote by {\bf S} and {\bf F}. This includes the littlest Higgs models. As in Ref.~\cite{top-hunter}, one can  consider two choices for the representation of the operator which induces the mixing of the elementary fermions with the strong sector, namely {\bf 5} and {\bf 14} of $SO(5)$. We then end up with four different choices of representations, top-partners in the singlet  (${\bf S_{5,14}}$) and fundamental (${\bf F_{5,14}}$) representations.

In Table~\ref{transl}, we show the translation between our parametrization and several benchmark models.
Using   
\bea
\xi=\frac{v^2}{f^2}\,=\,\sin^2{\epsilon}\,.
\eea
and $N=\sqrt{c^2_\epsilon+c^2_{2 \epsilon}}$,
\begin{table}[h!]
		\centering
		\setlength{\tabcolsep}{5pt}
		\begin{tabular}{|c|c|c|c|} 
		\hline 
	Model  & $m$  & $\Delta$  & $M$   	\\ \hline   \hline
	${\bf S_5}$ & $- \frac{c y\, f}{ \sqrt{2}} \sin{\epsilon}$ & $-\frac{ y\, f}{ \sqrt{2}} \sin{\epsilon}$  & $-M_{\Psi}$\\ \hline
	${\bf S_{14}}$ & $- \frac{c y\, f}{ 2\sqrt{2}} \sin{2 \epsilon}$ & $-\frac{ y\, f}{ 2\sqrt{2}} \sin{2 \epsilon}$ &  $-M_{\Psi}$\\ \hline
	${\bf F_{5}}$ & $- \frac{c y\, f}{ \sqrt{2}} \sin{\epsilon}$ & $y\,f{\sqrt{\cos^4{\frac{\epsilon}{2}}+\sin^4{\frac{\epsilon}{2}}}}$ & $-M_{\Psi}$\\ \hline
	${\bf F_{14}}$ & $- \frac{ y\,f}{ 2 \sqrt{2}} \sin{2\epsilon}$  &   $\frac{ y\,f}{2} cos{\epsilon} \, N$  & $-\, N^2 M_\Psi/4$ \\ \hline
		\end{tabular}
		\caption{\it Translation between our parametrization and the choices in the model space.}
	\label{transl}
\end{table}

\end{document}